\documentclass[aps,prd,preprint,superscriptaddress,amsmath,amssymb,showpacs]{revtex4-1}
\usepackage{dcolumn}
\usepackage{graphicx}
\usepackage{float}
\usepackage{subfloat}
\usepackage{amsmath}
\usepackage{physics}
\usepackage[colorlinks=true,allcolors=blue]{hyperref}
\usepackage{subcaption}
\usepackage[utf8]{inputenc}
\usepackage{array}
\usepackage{tabularx}
\usepackage{booktabs}

\begin{document}
\title{Holographic Schwinger effect in flavor-dependent systems}
\author{Sheng Lin}
\affiliation{School of Nuclear Science and Technology, University of South China, Hengyang 421001, China}
\author{Xuan Liu}
\affiliation{School of Nuclear Science and Technology, University of South China, Hengyang 421001, China}
\author{Xun Chen}
\email{chenxunhep@qq.com}
\affiliation{School of Nuclear Science and Technology, University of South China, Hengyang 421001, China}
\affiliation{Key Laboratory of Quark and Lepton Physics (MOE), Central China Normal University, Wuhan 430079,China}
\author{Gen-Fa Zhang}
\email{zhanggfa04@163.com}
\affiliation{School of Nuclear Science and Technology, University of South China, Hengyang 421001, China}
\author{Jing Zhou}
\email{zhoujing@hncu.edu.cn}
\affiliation{Department of Physics, Hunan City University, Yiyang, Hunan 413000, China}
\affiliation{All-solid-state Energy Storage Materials and Devices Key Laboratory of Hunan Province,
Hunan City University, Yiyang 413000, China}

\begin{abstract}
The holographic Schwinger effect is investigated in systems with $N_{f}=0$, $N_{f}=2$, and $N_{f}=2+1$ using the Einstein-Maxwell-dilaton (EMD) model, incorporating equation of state and baryon number susceptibility information from lattice quantum chromodynamics (QCD). It is found that the critical electric field is smallest for $N_{f}=0$, indicating that the Schwinger effect is more likely to occur than in systems with $N_{f}=2$ and $N_{f}=2+1$. The critical electric field decreases with increasing chemical potential and temperature across all systems. Additionally, potential analysis confirms that the maximum total potential energy increases with the number of flavors, suggesting that existing particles may reduce the probability of particle pair production.
\end{abstract}
\maketitle
	
\section{Introduction}\label{sec:01_intro}

The phenomenon of the Schwinger effect \cite{Schwinger:1951nm} elucidates the production of particle-antiparticle pairs from a vacuum under the influence of a strong electric field. This phenomenon stands as a crucial example of nonperturbative effects in quantum electrodynamics (QED) offering a unique window into the study of nonperturbative phenomena, as it cannot be explained through traditional perturbative methods. In high-energy particle colliders and heavy-ion collision experiments, studying particle generation under strong electric fields aids in exploring new physical phenomena. Schwinger's initial formulation for the production rate (per unit time and volume) in the regime of weak coupling and weak external electric fields (\(eE \leqslant m^{2}\)) is expressed as $\Gamma \sim \exp \left(-\frac{\pi m^2}{eE}\right)$. In conditions of arbitrary coupling and weak external electric fields, the production rate can be calculated using the Affleck-Alvarez-Manton (AAM) approach \cite{Affleck:1981bma}, yielding $\Gamma \sim \exp \left(-\frac{\pi m^2}{eE} + \frac{e^2}{4}\right)$.

However, the critical value derived does not conform to the weak-field condition, leaving the existence of such a critical value unverified.
In the context of string theory, the critical electric field is proportional to the string tension \cite{Fradkin:1985ys,Bachas:1992bh}. The AdS/CFT correspondence \cite{Maldacena:1997re,Witten:1998qj,Aharony:1999ti, Bachas:1992bh, Gubser:1998bc} provides a bridge between string theory on \(AdS_{5} \times S^{5}\) space and \(\mathcal{N}=4\) super Yang-Mills (SYM) theory, enabling the study of the Schwinger effect through holographic principle. This correspondence translates the strongly coupled gauge theory into a weakly coupled gravitational theory, offering a robust paradigm for probing strongly coupled quantum field theories. Within this framework, to simulate the \(\mathcal{N}=4\) SYM system coupled with a \(U(1)\) gauge field, the gauge group \(U(N + 1)\) is broken down to \(SU(N) \times U(1)\) via the Higgs mechanism. In the holographic Schwinger effect, the critical external electric field can be obtained by introducing D3-branes. Moreover, the introduction of D3-branes helps to avoid the issue of the produced particles having infinite mass. Specifically, the mass of the produced particles is typically assumed to be infinite, which significantly suppresses the pair production rate in the holographic Schwinger effect. However, by placing a probe D3-brane at a finite radial position $z_0$, rather than at the AdS boundary, the integral range can be limited, preventing the particle mass from becoming infinite \cite{Semenoff:2011ng}.

So far, numerous studies on the Schwinger effect have been carried out both in top-down and bottom-up approaches, significantly enhancing our understanding of this effect and the mechanisms of particle production. The study in Ref. \cite{Li:2022hka} delved into the phenomena of the Schwinger effect and electric instability within the backdrop of anisotropy in type IIB supergravity. In Ref. \cite{Ghoroku:2016kft}, they probed the instability induced by an external electric field using D7 probe branes in a type IIB theory framework. Besides, in Ref. \cite{Zhang:2020noe}, the authors examined the holographic Schwinger effect in a deformed AdS space with a gluon condensate, demonstrating that the presence of the condensate diminishes the particle production rate. In a rotating strongly coupled medium, it is found that the holographic Schwinger effect is augmented by the medium's angular velocity  \cite{Cai:2022nwq}. Many recent studies on the Schwinger effect with different contexts have been carried out in Refs. \cite{Cai:2023cjl,Zhou:2021nbp,Zhu:2021ucv,Ding:2020slu,Li:2020azb,Zhang:2017egb,Xu:2016cdt,Zhang:2016qiz,Sadeghi:2016ppy,Shahkarami:2015qff,Zhang:2015bha,Kawai:2015mha,Chakrabortty:2014kma,Sato:2013hyw,
Sato:2013dwa,Fischler:2014ama,BitaghsirFadafan:2015asm,Zhu:2024pdx,Wu:2015krf,Sato:2013pxa}.

Hydrodynamical models require an equation of state (EoS) as an input to relate local thermodynamic quantities. Therefore, an accurate determination of the quantum chromodynamics (QCD) EoS is crucial for understanding the nature of the matter produced in heavy ion collisions, as well as for modeling the behavior of hot matter in the early universe \cite{Borsanyi:2010cj}. Lattice QCD has studied EoS with different flavors in Refs. \cite{Borsanyi:2011wyg,Borsanyi:2012ve,Fischer:2018sdj,Ratti:2013uta,Weber:2021hro}. Quark (baryon) number susceptibilities have become important objects of study in recent years. They are important ingredients in the determination of the phase diagram of QCD as well as the equation of state (EoS) of strongly interacting matter \cite{Datta:2016ukp}. Recent lattice QCD has carried out many works on the quark (baryon) number susceptibilities for different flavors \cite{Bellwied:2015lba,Bazavov:2017dus,Sharma:2022ztl}.

In our recent works \cite{Chen:2024ckb,Chen:2024mmd}, we have incorporated the equation of state (EoS) and baryon number susceptibility for different flavors into the holographic model using machine learning technology. We further calculated the potential energy and transport properties of heavy quarks at finite temperature and chemical potential for systems with varying flavors \cite{Guo:2024qiq,Chen:2024epd}. These results are consistent with experimental data and other models, thereby confirming the validity of the holographic model.

Based on this model, our primary focus is on studying the Schwinger effect in the systems with various flavors in this paper. Specifically, we aim to ascertain the influence of different flavors on the total potential energy and determine whether the critical electric field and separation length are affected. The remainder of the paper is organized as follows. In Sec. \ref{sec:02}, we introduce the 5-dimensional Einstein-Maxwell-dilaton (EMD) systems. In Sec. \ref{sec:04}, we analyze the separation lengths in different flavor contexts and chemical potentials. Section \ref{sec:05} is devoted to discussing the total potential and addressing the critical field. Finally, discussions and conclusions are presented in Sec. \ref{sec:06}.

\section{The Setup}\label{sec:02}

\subsection{Background geometry}
References \cite{DeWolfe:2010he,DeWolfe:2011ts} opened the now very prolific research area on EMD models applied to QCD matter. The 5-dimensional EMD model has proved suitable for describing the transport properties \cite{DeWolfe:2011ts,Rougemont:2015wca,Rougemont:2017tlu,Grefa:2022sav,Chen:2024epd}, heavy quark potential \cite{Zhou:2020ssi,Guo:2024qiq}, the QCD phase diagram \cite{DeWolfe:2010he,He:2013qq,Yang:2014bqa,Yang:2015aia,Dudal:2018ztm,Chen:2018vty,Chen:2020ath,Chen:2019rez,Knaute:2017opk,Grefa:2021qvt,Cai:2022omk,Li:2023mpv,
Rougemont:2023gfz,Zhao:2023gur,Fu:2024wkn,Jokela:2024xgz,Cai:2024eqa,Chen:2024ckb,Chen:2024mmd}, etc. The action in the Einstein frame is given by
\begin{equation}\label{e1}
	S = \int \frac{d^{5}x}{16\pi G_{5}} \sqrt{-\text{det}(g_{\mu\nu})} \left[ R - \frac{f(\phi)}{4}F^{2}  - \frac{1}{2}\partial_{\mu}\phi\partial^{\mu}\phi - V(\phi) \right].
\end{equation}
Here, $F$ is Maxwell field with field strength tensors $F_{\mu\nu}^{(1)} = \partial_{\mu}A_{\nu} - \partial_{\nu}A_{\mu}$. The function $f(\phi)$ corresponds to the gauge function for the Maxwell field. $V(\phi)$ is the scalar potential, and $G_{5}$ is the Newton constant in five dimensions. The explicit forms of the gauge kinetic function $f(\phi)$ and the dilaton potential $V(\phi)$ can be solved consistently from the equations of motion.

The metric ansatz for the black brane solution in the system is given by
\begin{equation}\label{metric1}
	ds^{2} = \frac{L^{2}e^{2A(z)}}{z^{2}} \left[-g(z)dt^{2} + \frac{dz^{2}}{g(z)} + d\vec{x}^{2}\right],
\end{equation}
where $z$ is the 5th-dimensional holographic coordinate, and the AdS$_{5}$ space radius $L$ is set to one ($L = 1$). The boundary conditions near the horizon are
\begin{equation}
A_t\left(z_t\right)=g\left(z_t\right)=0.
\end{equation}
$z_t$ is the horizon of the black hole, which is related to the temperature. Near the boundary $z \rightarrow 0$, we require the metric in the string frame to be asymptotic to $\rm AdS_5$. The boundary conditions are
\begin{equation}
A(0)=-\sqrt{\frac{1}{6}} \phi(0), \quad g(0)=1, \quad \phi(0) = 0, \quad A_t(0)=\mu+\rho^{\prime} z^2+\cdots.
\end{equation}
$\mu$ can be regarded as baryon chemical potential and $\rho^{\prime}$ is proportional to the baryon number density. $\mu$ is related to the quark-number chemical potential $\mu = 3\mu_q$. Following the methodology in \cite{Chen:2024ckb,Chen:2024mmd}, we have following solutions
\begin{equation}
\begin{aligned}
g(z)&=1-\frac{1}{\int_0^{z_t} d x x^3 e^{-3 A(x)}}\left[\int_0^z d x x^3 e^{-3 A(x)}+\frac{2 c \mu^2 e^k}{\left(1-e^{-c z_t^2}\right)^2} \operatorname{det} \mathcal{G}\right],\\
\phi^{\prime}(z) & =\sqrt{6\left(A^{\prime 2}-A^{\prime \prime}-2 A^{\prime} / z\right)}, \\
A_t(z) & =\mu \frac{e^{-c z^2}-e^{-c z_t^2}}{1-e^{-c z_t^2}}, \\
V(z) & =-\frac{3 z^2 g e^{-2 A}}{L^2}\left[A^{\prime \prime}+A^{\prime}\left(3 A^{\prime}-\frac{6}{z}+\frac{3 g^{\prime}}{2 g}\right)-\frac{1}{z}\left(-\frac{4}{z}+\frac{3 g^{\prime}}{2 g}\right)+\frac{g^{\prime \prime}}{6 g}\right],
\end{aligned}
\end{equation}
where
\begin{equation}
\operatorname{det} \mathcal{G}=\left|\begin{array}{ll}
\int_0^{z_t} d y y^3 e^{-3 A(y)} & \int_0^{z_t} d y y^3 e^{-3 A(y)-c y^2} \\
\int_{z_t}^z d y y^3 e^{-3 A(y)} & \int_{z_t}^z d y y^3 e^{-3 A(y)-c y^2}
\end{array}\right|.
\end{equation}
The temperature can be given as
\begin{equation}
\begin{aligned} T=&-\frac{g(z_{t})^{'}}{4\pi}\\=&\frac{z_{t}^{3}e^{-3A(z_{t})}}{4\pi\int_{0}^{z_{t}}dyy^{3}e^{-3A(y)}}[1+\frac{2c\mu^{2}e^{k}(e^{-cz_{t}^{2}}\int_{0}^{z_{t}} dyy^{3}e^{-3A(y)}-\int_{0}^{z_{t}}dyy^{3}e^{-3A(y)-cy^{2}} )}{(1-e^{-cz^{2}_{t}})^{2}}].
	\end{aligned}
\end{equation}	
Given the functions $f(z)=e^{c z^2-A(z)+k}$ and $A(z)= d*\ln(a z^2 + 1) + d*\ln(b z^4 + 1)$, we can get the full solutions of the model. For convenience, we transform to the string frame in order to calculate the Nambu-Goto action later. Note that in the string frame, $A_{s}(z) = A(z) + \sqrt{1/6}\phi(z)$. By inputting the information of EoS and baryon number susceptibility from lattice QCD, we constructed the model for different flavors with six parameters as shown in Tab.~\ref{1} \cite{Chen:2024ckb,Chen:2024mmd}

\begin{table}[H]
	\centering
	\begin{tabular}{|l|c|c|c|c|c|c|}
		\hline
		& $a$ & $b$ & $c$ & $d$ & $k$ & $G_{5}$ \\
		\hline
		$N_{f}=0$ & 0 & 0.072 & 0 & -0.584 & 0 & 1.326 \\
		\hline
		$N_{f}=2$ & 0.067 & 0.023 & -0.377 & -0.382 & 0 & 0.885 \\
		\hline
		$N_{f}=2+1$ & 0.204 & 0.013 & -0.264 & -0.173 & -0.824 & 0.400 \\
		\hline
	\end{tabular}
	\caption{The parameters for the systems with various flavors are used in our model.}
	\label{1}
\end{table}

\begin{figure}
\centering
	\includegraphics[width=\linewidth]{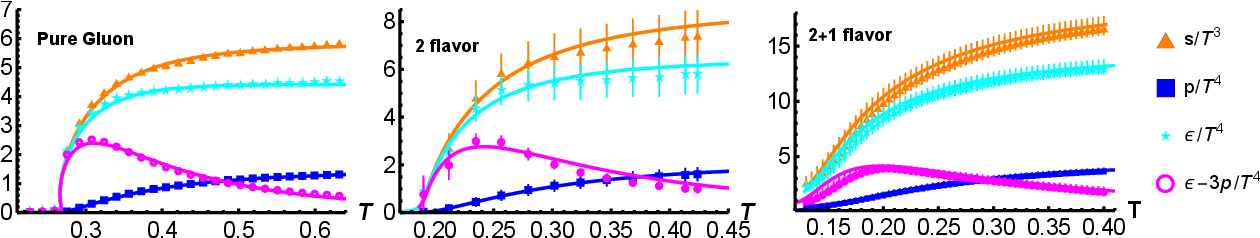}
	\caption{\label{eosall} Holographic EoS with lattice data for pure gluon, 2-flavor, and 2+1-flavor systems \cite{Chen:2024ckb}.}
\end{figure}

To be more clear, we show the EoS for the systems with different flavors calculated from holographic QCD and lattice QCD as shown in Fig.~\ref{eosall}.

\subsection{Potential analysis}\label{sec:03}
\begin{figure}
	\includegraphics[width=0.7\linewidth]{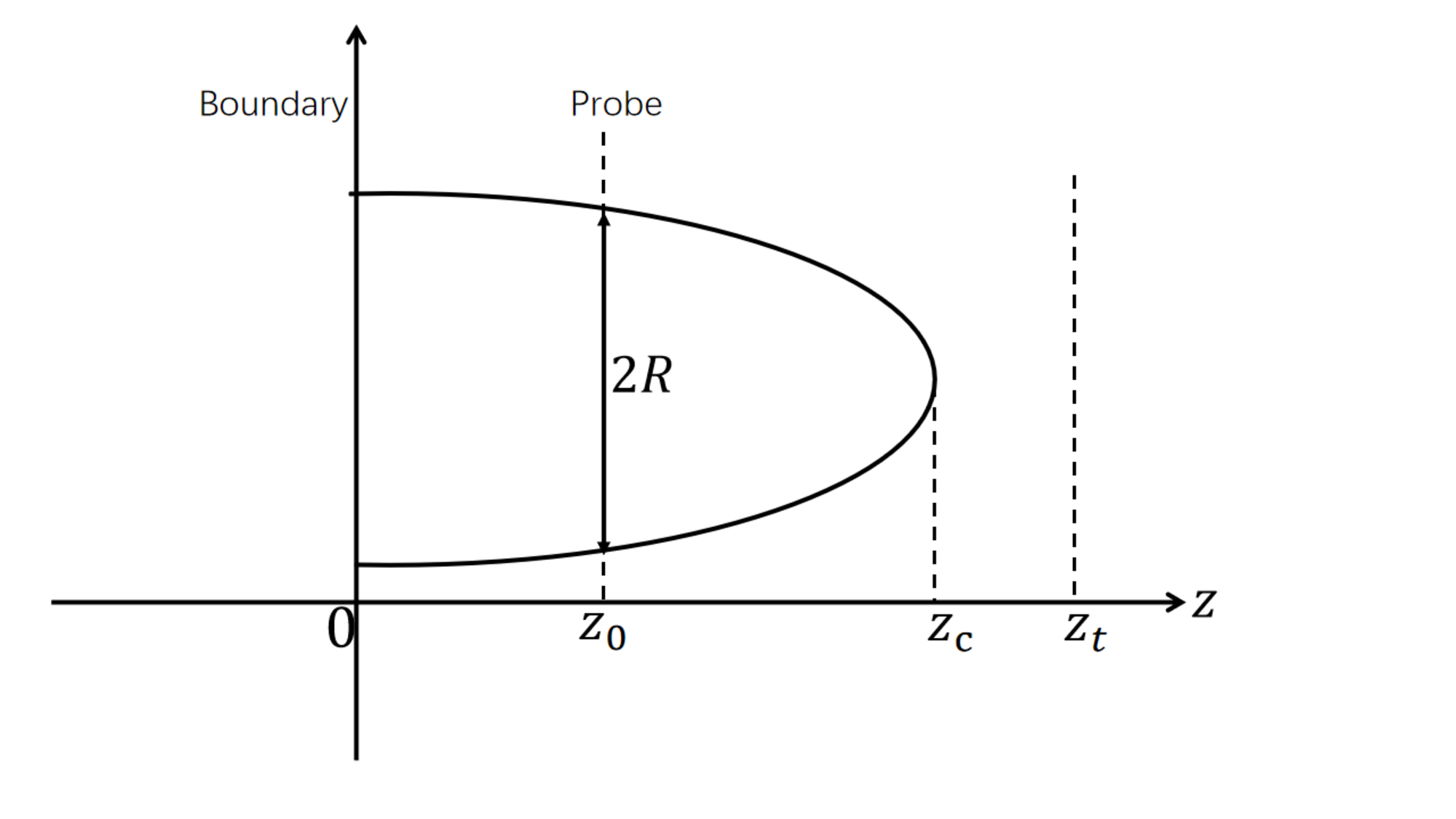}
	\caption{\label{Fa}  The configuration of the string world sheet for the Schwinger pair production.}
\end{figure}

The Schwinger effect refers to the process in which real particle pairs are produced from the vacuum under the influence of an external electric field \cite{Sato:2013hyw}. In the holographic Schwinger effect, the critical external electric field can be obtained by introducing D3-branes. Moreover, the introduction of D3-branes helps to avoid the issue of the produced particles having infinite mass. Specifically, the mass of the produced particles is typically assumed to be infinite, which significantly suppresses the pair production rate in the holographic Schwinger effect. However, by placing a probe D3-brane at a finite radial position $z_0$, rather than at the AdS boundary, the integral range can be limited, preventing the particle mass from becoming infinite. From Refs. \cite{Sato:2013hyw,Sato:2013dwa,Sato:2013iua,Kawai:2015mha,Li:2020azb,Zhou:2021nbp}, the DBI action of the probe D3 brane is
\begin{equation}
	S_{DBI}=-T_{D3}\int d^{4}x\sqrt{-det(G_{\mu\nu}+\mathcal{F_{\mu\nu}})}.
\end{equation}
Here, $\mu, \nu$ run through 0, 1, 2, 3. From the metric in Eq. (\ref{metric1}), we have $G_{00} = -g(z)\frac{b(z)}{z^{2}}$, $G_{11} = G_{22} = G_{33} = \frac{b(z)}{z^{2}}$. The D3-brane tension, denoted as $T_{D3}$, is given by the formula $T_{D3} = \frac{1}{g_s (2\pi)^3 \alpha^{\prime 2}}$, where the string tension is $T_F = \frac{1}{2\pi \alpha^{\prime}}$ and the string coupling constant is $g_s$ \cite{Sato:2013pxa}. Note the $\mathcal{F_{\mu\nu}}$ term can be written as $\mathcal{F_{\mu\nu}} = \alpha^{'}F_{\mu\nu}$. If we consider the external electric field to be oriented along the $x_1$ direction, then the expansion of $G_{\mu\nu}+\mathcal{F_{\mu\nu}}$ would be as follows
\begin{equation}
	G_{\mu\nu}+\mathcal{F_{\mu\nu}}=\begin{bmatrix}
		-g(z)\frac{b(z)}{z^{2}} & 2\pi\alpha^{'}E & 0 & 0 \\
		-2\pi\alpha^{'}E  & \frac{b(z)}{z^{2}}  & 0 & 0 \\
		0 & 0 & \frac{b(z)}{z^{2}} & 0\\
		0 & 0 & 0 & \frac{b(z)}{z^{2}} \\	
	\end{bmatrix},
\end{equation}
where $b(z) = e^{2As}$. We can drive the DBI action at $z = z_0$ as
\begin{equation}
	S_{DBI}=-T_{D3}\int d^{4}x\frac{b(z_0)}{z^4}\sqrt{-b\left(z_{0}\right)^{2}g\left(z_{0}\right)+\left(2\pi\alpha^{\prime}E\right)^{2}z_{0}^{4}}.
\end{equation}
$z=z_{0}$ is the probe D3 brane's position, which is related with the mass of the particle \cite{Sato:2013pxa,Sato:2013iua, Shahkarami:2015qff}. We set $z_0=0.6$ $\rm GeV^{-1}$ to investigate the qualitative behavior of Schwinger effect in our case. The physical interpretation of the region between the boundary and the flavor D3-brane in Fig. \ref{Fa} can be explained as follows: First, it is important to note that \( z_0 \) is not fixed and can vary between 0 and \( z_c \). This means that the distance between the boundary and the D3-brane ranges from 0 to \( z_c \). By positioning the D3-brane at \( z_0 \), we are able to calculate the separation distance between the particle-antiparticle pairs. Changing the value of \( z_0 \) does not alter the nature of the separation distance but only affects its numerical value. If the equation is to have physical significance, it becomes necessary to require that the second radical be greater than or equal to zero \cite{Sato:2013hyw,Sato:2013dwa,Sato:2013iua,Kawai:2015mha,Li:2020azb,Zhou:2021nbp}, namely,
\begin{equation}
	-b\left(z_{0}\right)^{2}g\left(z_{0}\right)+\left(2\pi\alpha^{\prime}E\right)^{2}z_{0}^{4} \geqslant 0.
\end{equation}
So the critical electric field can be determined as
\begin{equation}
	E_{c}=T_{F}\sqrt{\frac{b(z_{0})^{2}g(z_{0})}{z_{0}^{4}}}.
\end{equation}
We turn to the Nambu-Goto action action of string on the probe D3-brane. It is convenient to choose the world sheet coordinates to be
\begin{align}
	x_{0} = \tau, ~ x_{1} = \sigma, ~ x_{2} = 0,~  x_{3} = 0,~  z = z(\sigma).
\end{align}
Then, we have
\begin{equation}
ds^2 = \frac{b(z)}{z^2}\left(-g(z) d\tau^2 + \frac{1}{g(z)} \left(\frac{dz}{d\sigma}\right)^2 d\sigma^2 + d\sigma^2\right).
\end{equation}
The Nambu-Goto action is given by
\begin{equation}
	S_{NG} = - \frac{1}{2\pi\alpha'} \int d\tau d\sigma \sqrt{- \det g_{ab}}.
\end{equation}
Here, $g_{ab}$ is the induced metric on the world sheet with $a, b = 0, 1$, corresponding to the coordinates $\tau$ and $\sigma$. It follows that $g_{00} = -\frac{b(z)g(z)}{z^2}$ and $g_{11} = -\frac{b(z)}{z^2}\left(\frac{\dot{z}^2}{g(z)} + 1\right)$. Here, we have defined $\dot{z} = \frac{dz}{d\sigma}$ for simplicity. $\frac{1}{2\pi\alpha'}$ is related to the fundamental string tension $T_F$. We can write down the Lagrangian
\begin{equation}
	\mathcal{L} = \frac{b(z)}{z^{2}}\sqrt{g(z) + \dot{z}^{2}}.
\end{equation}
Note that Lagrangian does not explicitly depend on $x$, which represents the separation length of the particle pair. Thus, we have the conserved quantity
\begin{equation}
	\mathcal{L} - \frac{\partial \mathcal{L}}{\partial \dot{z}}\dot{z} = \mathcal{C}.	\label{10}
\end{equation}
Besides, we have the following boundary conditions as shown in Fig. \ref{Fa}
\begin{align}
	\frac{dz}{d\sigma} =0, \quad z =z_{c} \quad (z_{0}<z_{c}<z_{t}).
\end{align}
Thus, we can get
\begin{equation}
	\mathcal{C}=\frac{b(z_{c})}{z_{c}^{2}}\sqrt{g(z_{c})}. \label{11}
\end{equation}
From Eq. (\ref{10}) and Eq. (\ref{11}), we have
\begin{equation}
	\dot{z}=\sqrt{(\frac{b(z)g(z)}{z^{2}\mathcal{C}})^{2}-g(z)}.
\end{equation}
We can obtain the separation length of the particle pair
\begin{equation}\label{17}
	x=2R=2\int_{z_{0}}^{z_{c}}dz\frac{1}{\sqrt{(\frac{b(z)g(z)}{z^{2}\mathcal{C}})^{2}-g(z)}}.
\end{equation}
The Schwinger effect refers to the process where virtual particle pairs in a vacuum gain energy from an external field and become real particles. The amount of energy required for this transition can be analyzed through the potential energy. Specifically, for virtual particles to become real, the potential energy required is given by \cite{Sato:2013hyw}
\begin{eqnarray}
V_{\text {tot }}(x) & = & 2 m-\frac{\alpha_{\mathrm{s}}}{x}-E x
\end{eqnarray}
The first term on the right side of the equation represents the static energy (SE), the second term is the Coulomb potential (CP), and the third term corresponds to the energy from the external electric field. The sum of the Coulomb potential and the static energy is \cite{Sato:2013hyw}
\begin{eqnarray}
V_{(C P+S E)} =  2 T_{F} \int_{0}^{x / 2} d x \mathcal{L}  = 2 T_{F} \int_{z_{c}}^{z_{0}} d z \frac{b(z)^{2} g(z)}{z^{2} \sqrt{(b(z) g(z))^{2}-z^{4} g(z) \mathcal{C}^{2}}}
\end{eqnarray}
By adding the energy from the external field and defining $\beta= \frac{E}{E_{c}}$, the total potential energy becomes
\begin{eqnarray}
V_{\text {tot }}  & = & V_{(C P+S E)}-E x \\
 & = & 2 T_{F}\left(\int_{z_{0}}^{z_{c}} d z \frac{b(z)^{2} g(z)}{z^{2} \sqrt{(b(z) g(z))^{2}-z^{4} g(z) \mathcal{C}^{2}}}-\beta \int_{z_{0}}^{z_{c}} d z \frac{\sqrt{\frac{b\left(z_{0}\right)^{2} g\left(z_{0}\right)}{z_{0}^{4}}}}{\sqrt{(\frac{b(z)g(z)}{z^{2}\mathcal{C}})^{2}-g(z)}}\right),
\end{eqnarray}
where $x$ is given by Eq.~(\ref{17}).

\section{The separation length x in different flavors}\label{sec:04}

\begin{figure}
		\centering
	\begin{subfigure}[b]{0.32\linewidth}
		\includegraphics[width=\linewidth]{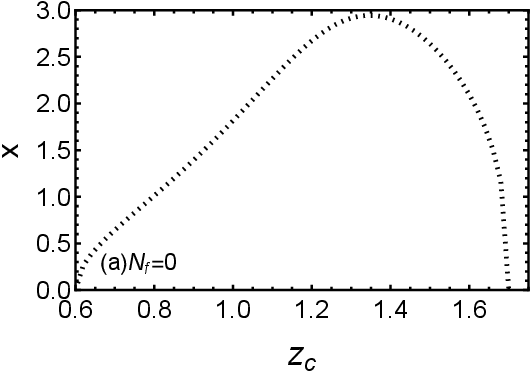}
		\label{fig:sub1}
	\end{subfigure}
	\hfill
	\begin{subfigure}[b]{0.32\linewidth}
		\includegraphics[width=\linewidth]{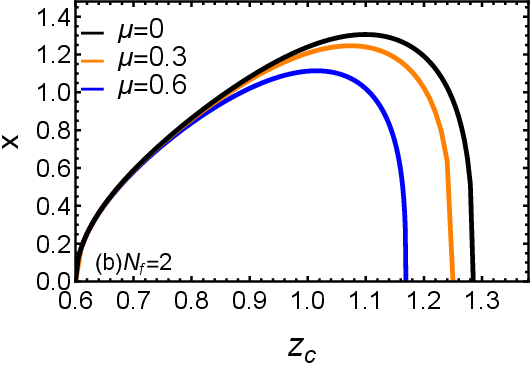}
		\label{fig:sub2}
	\end{subfigure}
	\hfill
	\begin{subfigure}[b]{0.32\linewidth}
		\includegraphics[width=\linewidth]{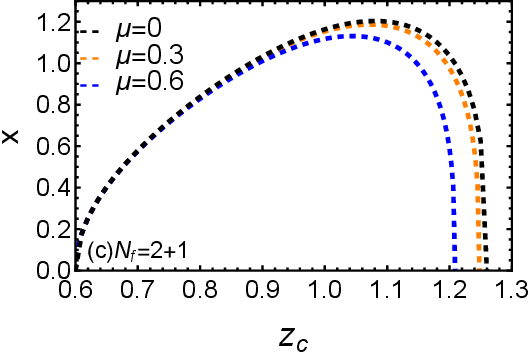}
		\label{fig:sub3}
	\end{subfigure}
	\caption{\label{F2} (a) The separation length $x$ versus $z_{c}$ at $N_{f} = 0$ and $T$ = 0.266. (b) The separation length $x$ versus the parameter $z_{c}$ at $N_{f} = 2$. (c) The separation length $x$ versus the parameter $z_{c}$ at $N_{f} = 2+1$. Black, orange, blue represent $\mu = 0$, $\mu = 0.3$, $\mu = 0.6$. The units of $x$ and $z_{c}$ are GeV$^{-1}$. The units of $T$ and $\mu$ are GeV.}
	\label{fig:main}
\end{figure}

To investigate how chemical potential influences the separation distance $x$ in different systems, we present the results in Fig.~\ref{F2}. First, for simplicity, we set $T_{F}$ as constants and the external electric field strength to 3.5 GeV. Second, as illustrated in Fig.~\ref{F2}, we observe that the maximum separation distance decreases with increasing chemical potential for both $N_{f} = 2$ and $N_{f} = 2+1$. As the number of flavors in the system increases, a shorter separation distance is required to reach the same $z_0$.  The maximum separation length here refers to the maximum length of the string connecting the particle-antiparticle pair produced by the Schwinger effect. Specifically, in this model, the particles are connected by a string. Since the integration range is finite, this value is finite and has an upper limit, known as the maximum separation distance. This maximum value implies that if the separation exceeds this limit, the string dissolve, meaning that the particle-antiparticle pair no longer exists.

\begin{figure}
	\includegraphics[width=0.4\linewidth]{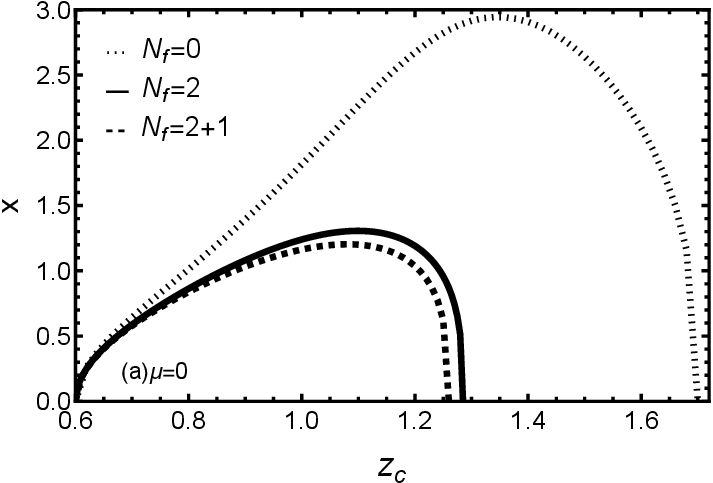}
	\caption{\label{2} The separation length $x$ versus the parameter $z_{c}$ at $\mu = 0$ and $T$ = 0.266. The units of $T$ and $\mu$ are GeV. }
\end{figure}

In the following, we aim to clarify how separation distances vary across different flavor systems, with the calculation results presented in Fig.~\ref{2}. The results show that the maximum separation distance is the greatest for $N_f = 0$, slightly less for $N_f = 2$, and the least for $N_f = 2+1$. This suggests that the addition of $u(d)$ quarks lead to a reduction in the maximum separation distance, and a further decrease is observed with the inclusion of an $s$ quark. In order to determine the effects of different systems on the Schwinger effect, we will calculate the total potential in the following section.

\section{The total potential in different flavors}\label{sec:05}
\begin{figure}
	\centering
    \begin{subfigure}[b]{0.32\linewidth}
		\includegraphics[width=\linewidth]{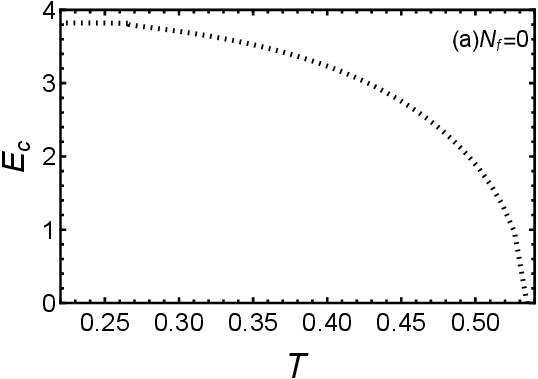}
		\label{fig:sub1}
	\end{subfigure}
	\hfill 
	\begin{subfigure}[b]{0.32\linewidth}
		\includegraphics[width=\linewidth]{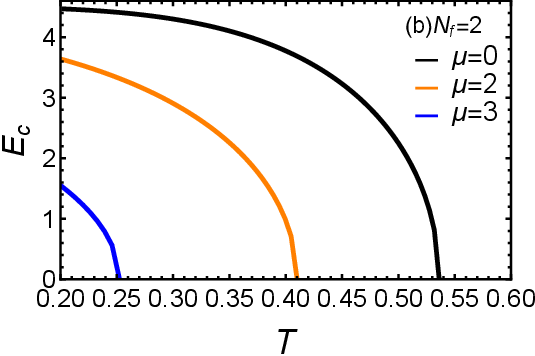}
		\label{fig:sub2}
	\end{subfigure}
	\hfill 
	\begin{subfigure}[b]{0.32\linewidth}
		\includegraphics[width=\linewidth]{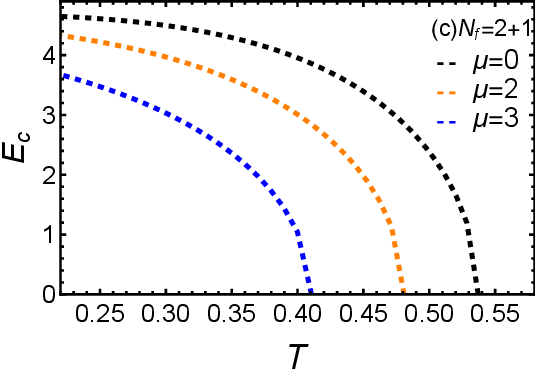}
		\label{fig:sub3}
	\end{subfigure}
\caption{\label{Fig4} The critical electric fields versus temperature under different flavors.}
\end{figure}

First, we investigate how temperature and chemical potential affect the critical electric field in different systems, as illustrated in Fig.~\ref{Fig4}. Figure~\ref{Fig4} shows that the critical electric field is smaller for the pure gluon system compared to the 2 flavor 2+1 flavor systems. We also observe that under the same temperature conditions, the critical electric field gradually decreases with increasing chemical potential across all flavor cases. This suggests that an increase in chemical potential enhances the possibility of the Schwinger effect, which aligns with previous findings \cite{Zhang:2018hfd}. Furthermore, as the temperature rises, the critical electric field is seem to decrease gradually, which indicates that higher temperature facilitate the production of particle pairs. This observation is consistent with the studies of \cite{Sato:2013iua, Zhu:2019igg, Zhou:2021nbp}.

\begin{figure}
\centering
	\begin{subfigure}[b]{0.32\linewidth}
		\includegraphics[width=\linewidth]{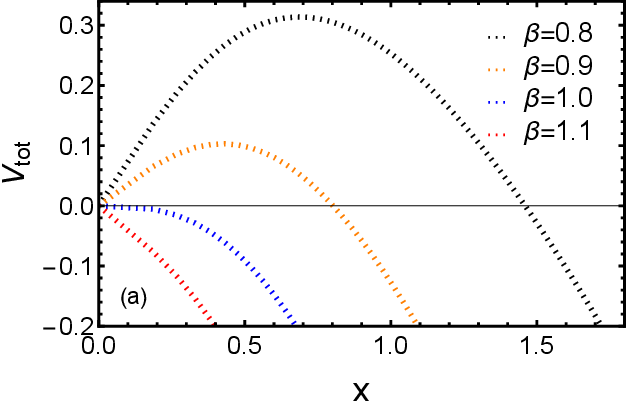}
		\label{fig:sub1}
	\end{subfigure}
	\hfill 
	\begin{subfigure}[b]{0.32\linewidth}
		\includegraphics[width=\linewidth]{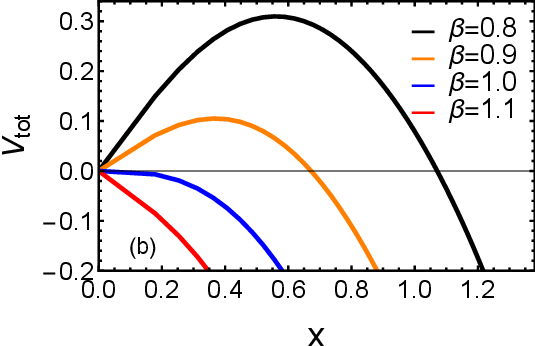}
		\label{fig:sub2}
	\end{subfigure}
	\hfill 
	\begin{subfigure}[b]{0.32\linewidth}
		\includegraphics[width=\linewidth]{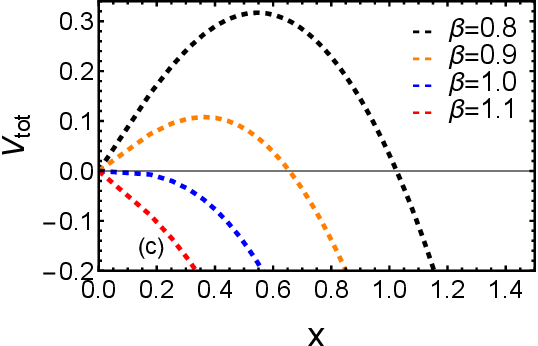}
		\label{fig:sub3}
	\end{subfigure}
	\caption{\label{Fig:5} (a) The total potential of $N_{f} =  0$ versus the separation length $x$. (b) The total potential of $N_{f} = 2$ versus the separation length $x$. (c) The total potential of $N_{f} = 2+1$ versus the separation length $x$. The temperature is fixed as $T = 0.266$ and the chemical potential is $\mu = 0$. The unit of $x$ is GeV$^{-1}$. The units of $T$, $V_{tot}$ and $\mu$ are GeV. Red, blue, orange, black lines are $\beta$=1.1, 1, 0.9, 0.8, respectively. }
	\label{fig:main}
\end{figure}

The critical electric fields, as calculated by the DBI action, are 3.80 GeV for $N_f = 0$, 4.38 GeV for $N_f = 2$ and 4.58 GeV for $N_f = 2+1$ at a fixed temperature of T = 0.266 GeV. This indicates that Schwinger pairs are more likely to occur for $N_f = 0$ and are less likely for $N_f = 2+1$. The potential barrier exists for $E < E_{c}$ (when $\beta < 1$), as shown in Fig.~\ref{Fig:5}. At $\beta = 1$, the barrier vanishes. Within this framework, particle production is conceptualized as a tunneling phenomenon. As the electric field progressively exceeds the critical threshold, the vacuum destabilizes, leading to the disappearance of the potential barrier, as described by \cite{Sato:2013iua}. Furthermore, Fig.~\ref{Fig:5} indicates that the barriers decrease as the external electric fields increase for systems with $N_{f} = 0$, $N_{f} = 2$, and $N_{f} = 2+1$, suggesting that particle generation becomes more probable. This is attributed to the virtual particle pairs gaining more energy from the external electric field.

\begin{figure}
\centering
		\begin{subfigure}[b]{0.32\linewidth}
		\includegraphics[width=\linewidth]{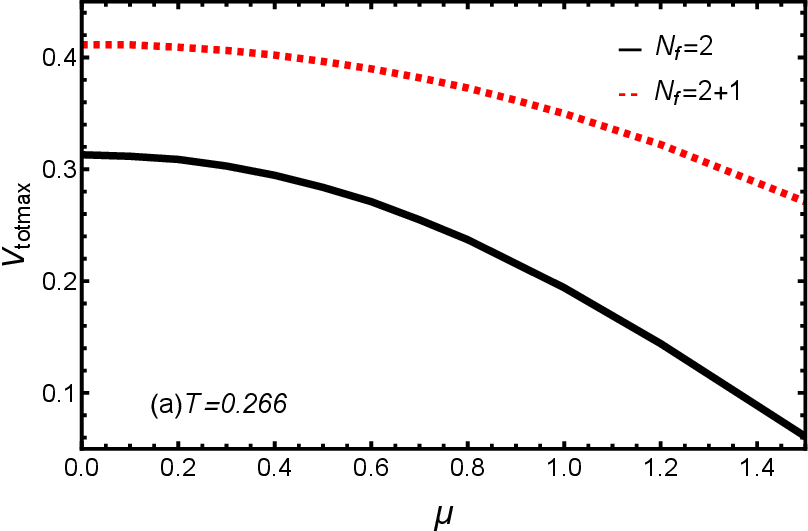}
		\label{fig:sub1}
	\end{subfigure}
	\hfill 
	\begin{subfigure}[b]{0.32\linewidth}
		\includegraphics[width=\linewidth]{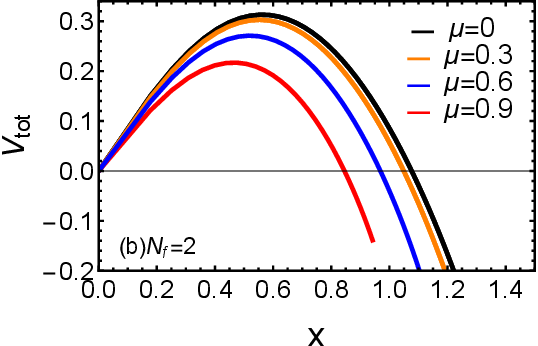}
		\label{fig:sub2}
	\end{subfigure}
	\hfill 
	\begin{subfigure}[b]{0.32\linewidth}
		\includegraphics[width=\linewidth]{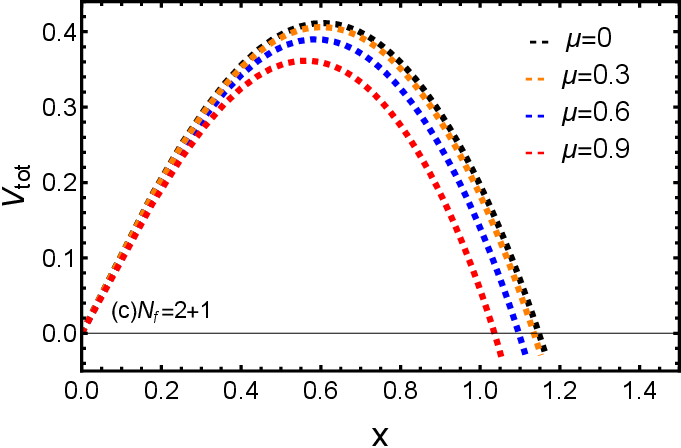}
		\label{fig:sub3}
	\end{subfigure}
	\caption{\label{Fig6}(a) The max total potentials change with $\mu$ at fixed $T = 0.266$. Red and black lines are $N_{f}$ = 2 and 2+1, respectively.  (b) The total potential of $N_{f}=2$ versus the separation length $x$ at fixed temperature $T = 0.266$. (c) The total potential of $N_{f}=2+1$ versus the separation length $x$ at fixed temperature $T=0.266$. We set external electric field $E$ as 3.5 GeV.  The unit of $x$ is GeV$^{-1}$. The units of $T$, $\mu$ and $V_{tot}$ are GeV. }
\end{figure}

\begin{figure}
\centering
	\includegraphics[width=\linewidth]{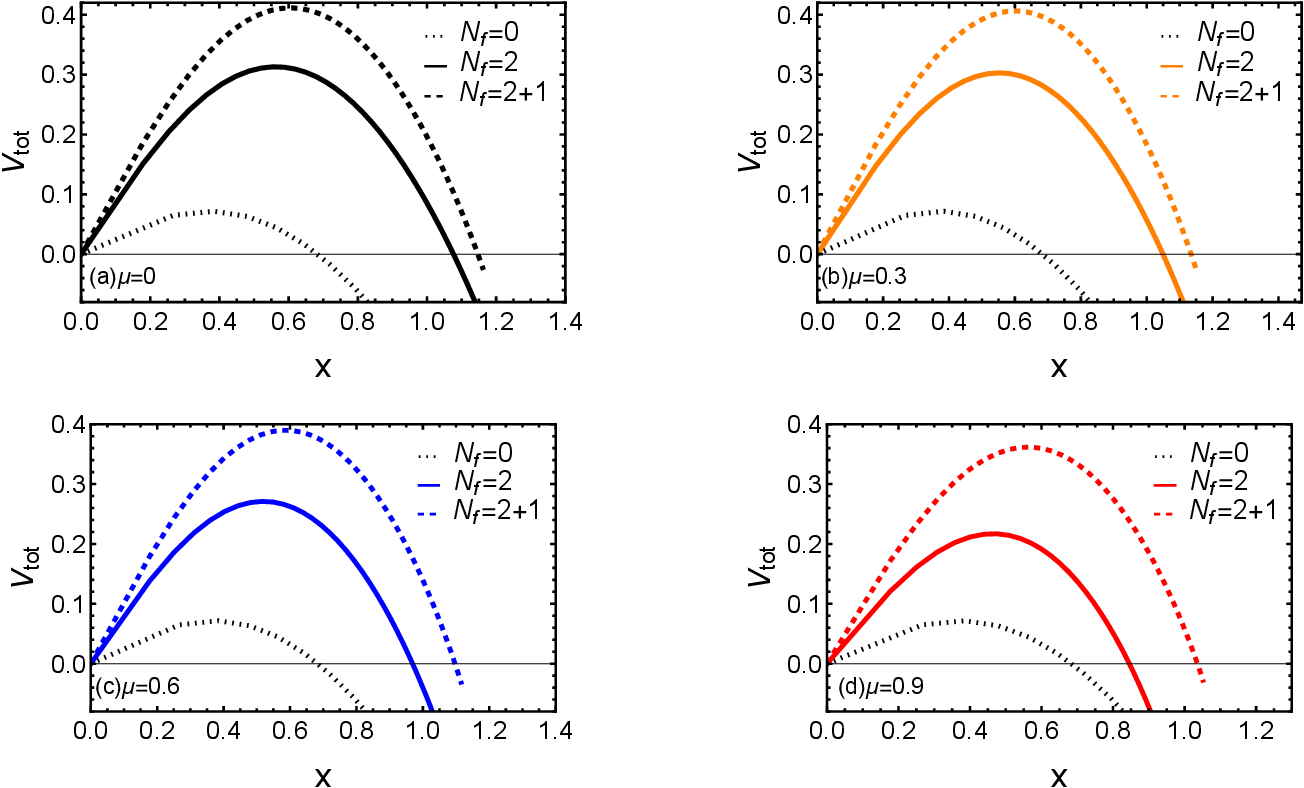}
	\caption{\label{Fig7}(a) The total potential versus the separation length $x$ at $\mu = 0$. (b) The total potential versus the separation length $x$ at $\mu = 0.3$. (c) The total potential versus the separation length $x$ at $\mu = 0.6$. (d) The total potential versus the separation length $x$ at $\mu = 0.9$. The dotted lines, full lines, dashed lines represent $N_{f} = 0$, $N_{f} = 2$, $N_{f} = 2+1$, respectively. We set external electric field $E$ as 3.5 GeV and temperature $T = 0.266$. Units for $x$ are GeV$^{-1}$, and units for $T$, $\mu$, and $V_{tot}$ are GeV.}
\end{figure}

To further investigate how chemical potential and flavor affect the total potential energy, we present the results in Figs.~\ref{Fig6} and ~\ref{Fig7}. Specifically, in Fig.~\ref{Fig6} (a), we observe that the maximum total potential energy increases as the number of flavors increases, suggesting that existing particles may reduce the likelihood of particle pair production. Furthermore, across Figs.~\ref{Fig6} (a), (b), and (c), we note that the total potential energy decreases with increasing chemical potential, a trend that aligns with the findings reported in Ref. \cite{Zhang:2018hfd}.

As shown in Fig.~\ref{Fig7}, for $N_{f} = 0$, the potential is independent of the chemical potential $\mu$. Notably, when $\mu = 0$, the case of $N_{f} = 2+1$ has the highest potential, and the potential for $N_{f} = 2$ is greater than that for $N_{f} = 0$. As the chemical potential increases gradually, the potential for $N_{f} = 0$ remains constant, while it decreases for both $N_{f} = 2$ and $N_{f} = 2+1$. Additionally, the width of the potential barrier decreases more rapidly in the $N_f = 2$ case than in the $N_f = 2+1$ case. This decrease is more significant in the $N_f = 2$ scenario as $\mu$ increases. Consequently, potential barriers in systems with $N_{f} = 2$ exhibit reduced sensitivity to changes in the chemical potential.

\section{Discussion and Conclusion}\label{sec:06}
The Schwinger effect was initially described for particle-antiparticle pair production in QED and has since been extended to quark-antiquark pair production in QCD, reflecting an evolving understanding of the effect. In this work, we put the probe D3 brane into three thermodynamic backgrounds \( N_f = 0 \), \( N_f = 2 \), and \( N_f = 2+1 \) to investigate the Schwinger effect. We delve into the investigation of separation length, total potential, and critical electric field at finite temperatures and chemical potentials across various flavors. It is observed that the increasing number of flavors leads to a reduction in separation length and an increase in total potential and critical electric field. Notably, systems with more flavors show reduced sensitivity to decrements in chemical potential. The production of pairs through the Schwinger mechanism may be achievable with more moderate chromoelectric fields \cite{Gelis:2015kya}. This phenomenon could significantly contribute to our understanding of particle production in heavy ion collisions. Consequently, if a strong enough electric field is generated in heavy ion collisions, it provides an opportunity to investigate the Schwinger mechanism within such collisions. On the theoretical side, our findings could provide valuable insights into the study of the Schwinger mechanism, particularly in the context of heavy ion collisions rather than in a vacuum.

\section{acknowledgements}
This work is supported by National Natural Science Foundation of China (NSFC) Grants No. 12405154, the Natural Science Foundation of Hunan Province of China under Grants No. 2024JJ6108, the Research Foundation of Education Bureau of Hunan Province, China under Grant No. 22B0788, and Open Fund for Key Laboratories of the Ministry of Education under Grants No. QLPL2024P01.

\section*{DATA AVAILABILITY}

No data were created or analyzed in this study.

\section*{References}

\bibliography{ref}

\begin{thebibliography}{73}%
\makeatletter
\providecommand \@ifxundefined [1]{%
 \@ifx{#1\undefined}
}%
\providecommand \@ifnum [1]{%
 \ifnum #1\expandafter \@firstoftwo
 \else \expandafter \@secondoftwo
 \fi
}%
\providecommand \@ifx [1]{%
 \ifx #1\expandafter \@firstoftwo
 \else \expandafter \@secondoftwo
 \fi
}%
\providecommand \natexlab [1]{#1}%
\providecommand \enquote  [1]{``#1''}%
\providecommand \bibnamefont  [1]{#1}%
\providecommand \bibfnamefont [1]{#1}%
\providecommand \citenamefont [1]{#1}%
\providecommand \href@noop [0]{\@secondoftwo}%
\providecommand \href [0]{\begingroup \@sanitize@url \@href}%
\providecommand \@href[1]{\@@startlink{#1}\@@href}%
\providecommand \@@href[1]{\endgroup#1\@@endlink}%
\providecommand \@sanitize@url [0]{\catcode `\\12\catcode `\$12\catcode
  `\&12\catcode `\#12\catcode `\^12\catcode `\_12\catcode `\%12\relax}%
\providecommand \@@startlink[1]{}%
\providecommand \@@endlink[0]{}%
\providecommand \url  [0]{\begingroup\@sanitize@url \@url }%
\providecommand \@url [1]{\endgroup\@href {#1}{\urlprefix }}%
\providecommand \urlprefix  [0]{URL }%
\providecommand \Eprint [0]{\href }%
\providecommand \doibase [0]{http://dx.doi.org/}%
\providecommand \selectlanguage [0]{\@gobble}%
\providecommand \bibinfo  [0]{\@secondoftwo}%
\providecommand \bibfield  [0]{\@secondoftwo}%
\providecommand \translation [1]{[#1]}%
\providecommand \BibitemOpen [0]{}%
\providecommand \bibitemStop [0]{}%
\providecommand \bibitemNoStop [0]{.\EOS\space}%
\providecommand \EOS [0]{\spacefactor3000\relax}%
\providecommand \BibitemShut  [1]{\csname bibitem#1\endcsname}%
\let\auto@bib@innerbib\@empty
\bibitem [{\citenamefont {Schwinger}(1951)}]{Schwinger:1951nm}%
  \BibitemOpen
  \bibfield  {author} {\bibinfo {author} {\bibfnamefont {J.~S.}\ \bibnamefont
  {Schwinger}},\ }\href {\doibase 10.1103/PhysRev.82.664} {\bibfield  {journal}
  {\bibinfo  {journal} {Phys. Rev.}\ }\textbf {\bibinfo {volume} {82}},\
  \bibinfo {pages} {664} (\bibinfo {year} {1951})}\BibitemShut {NoStop}%
\bibitem [{\citenamefont {Affleck}\ \emph {et~al.}(1982)\citenamefont
  {Affleck}, \citenamefont {Alvarez},\ and\ \citenamefont
  {Manton}}]{Affleck:1981bma}%
  \BibitemOpen
  \bibfield  {author} {\bibinfo {author} {\bibfnamefont {I.~K.}\ \bibnamefont
  {Affleck}}, \bibinfo {author} {\bibfnamefont {O.}~\bibnamefont {Alvarez}}, \
  and\ \bibinfo {author} {\bibfnamefont {N.~S.}\ \bibnamefont {Manton}},\
  }\href {\doibase 10.1016/0550-3213(82)90455-2} {\bibfield  {journal}
  {\bibinfo  {journal} {Nucl. Phys. B}\ }\textbf {\bibinfo {volume} {197}},\
  \bibinfo {pages} {509} (\bibinfo {year} {1982})}\BibitemShut {NoStop}%
\bibitem [{\citenamefont {Fradkin}\ and\ \citenamefont
  {Tseytlin}(1985)}]{Fradkin:1985ys}%
  \BibitemOpen
  \bibfield  {author} {\bibinfo {author} {\bibfnamefont {E.~S.}\ \bibnamefont
  {Fradkin}}\ and\ \bibinfo {author} {\bibfnamefont {A.~A.}\ \bibnamefont
  {Tseytlin}},\ }\href {\doibase 10.1016/0550-3213(85)90559-0} {\bibfield
  {journal} {\bibinfo  {journal} {Nucl. Phys. B}\ }\textbf {\bibinfo {volume}
  {261}},\ \bibinfo {pages} {1} (\bibinfo {year} {1985})},\ \bibinfo {note}
  {[Erratum: Nucl.Phys.B 269, 745--745 (1986)]}\BibitemShut {NoStop}%
\bibitem [{\citenamefont {Bachas}\ and\ \citenamefont
  {Porrati}(1992)}]{Bachas:1992bh}%
  \BibitemOpen
  \bibfield  {author} {\bibinfo {author} {\bibfnamefont {C.}~\bibnamefont
  {Bachas}}\ and\ \bibinfo {author} {\bibfnamefont {M.}~\bibnamefont
  {Porrati}},\ }\href {\doibase 10.1016/0370-2693(92)90806-F} {\bibfield
  {journal} {\bibinfo  {journal} {Phys. Lett. B}\ }\textbf {\bibinfo {volume}
  {296}},\ \bibinfo {pages} {77} (\bibinfo {year} {1992})},\ \Eprint
  {http://arxiv.org/abs/hep-th/9209032} {arXiv:hep-th/9209032} \BibitemShut
  {NoStop}%
\bibitem [{\citenamefont {Maldacena}(1998)}]{Maldacena:1997re}%
  \BibitemOpen
  \bibfield  {author} {\bibinfo {author} {\bibfnamefont {J.~M.}\ \bibnamefont
  {Maldacena}},\ }\href {\doibase 10.4310/ATMP.1998.v2.n2.a1} {\bibfield
  {journal} {\bibinfo  {journal} {Adv. Theor. Math. Phys.}\ }\textbf {\bibinfo
  {volume} {2}},\ \bibinfo {pages} {231} (\bibinfo {year} {1998})},\ \Eprint
  {http://arxiv.org/abs/hep-th/9711200} {arXiv:hep-th/9711200} \BibitemShut
  {NoStop}%
\bibitem [{\citenamefont {Witten}(1998)}]{Witten:1998qj}%
  \BibitemOpen
  \bibfield  {author} {\bibinfo {author} {\bibfnamefont {E.}~\bibnamefont
  {Witten}},\ }\href {\doibase 10.4310/ATMP.1998.v2.n2.a2} {\bibfield
  {journal} {\bibinfo  {journal} {Adv. Theor. Math. Phys.}\ }\textbf {\bibinfo
  {volume} {2}},\ \bibinfo {pages} {253} (\bibinfo {year} {1998})},\ \Eprint
  {http://arxiv.org/abs/hep-th/9802150} {arXiv:hep-th/9802150} \BibitemShut
  {NoStop}%
\bibitem [{\citenamefont {Aharony}\ \emph {et~al.}(2000)\citenamefont
  {Aharony}, \citenamefont {Gubser}, \citenamefont {Maldacena}, \citenamefont
  {Ooguri},\ and\ \citenamefont {Oz}}]{Aharony:1999ti}%
  \BibitemOpen
  \bibfield  {author} {\bibinfo {author} {\bibfnamefont {O.}~\bibnamefont
  {Aharony}}, \bibinfo {author} {\bibfnamefont {S.~S.}\ \bibnamefont {Gubser}},
  \bibinfo {author} {\bibfnamefont {J.~M.}\ \bibnamefont {Maldacena}}, \bibinfo
  {author} {\bibfnamefont {H.}~\bibnamefont {Ooguri}}, \ and\ \bibinfo {author}
  {\bibfnamefont {Y.}~\bibnamefont {Oz}},\ }\href {\doibase
  10.1016/S0370-1573(99)00083-6} {\bibfield  {journal} {\bibinfo  {journal}
  {Phys. Rept.}\ }\textbf {\bibinfo {volume} {323}},\ \bibinfo {pages} {183}
  (\bibinfo {year} {2000})},\ \Eprint {http://arxiv.org/abs/hep-th/9905111}
  {arXiv:hep-th/9905111} \BibitemShut {NoStop}%
\bibitem [{\citenamefont {Gubser}\ \emph {et~al.}(1998)\citenamefont {Gubser},
  \citenamefont {Klebanov},\ and\ \citenamefont {Polyakov}}]{Gubser:1998bc}%
  \BibitemOpen
  \bibfield  {author} {\bibinfo {author} {\bibfnamefont {S.~S.}\ \bibnamefont
  {Gubser}}, \bibinfo {author} {\bibfnamefont {I.~R.}\ \bibnamefont
  {Klebanov}}, \ and\ \bibinfo {author} {\bibfnamefont {A.~M.}\ \bibnamefont
  {Polyakov}},\ }\href {\doibase 10.1016/S0370-2693(98)00377-3} {\bibfield
  {journal} {\bibinfo  {journal} {Phys. Lett. B}\ }\textbf {\bibinfo {volume}
  {428}},\ \bibinfo {pages} {105} (\bibinfo {year} {1998})},\ \Eprint
  {http://arxiv.org/abs/hep-th/9802109} {arXiv:hep-th/9802109} \BibitemShut
  {NoStop}%
\bibitem [{\citenamefont {Semenoff}\ and\ \citenamefont
  {Zarembo}(2011)}]{Semenoff:2011ng}%
  \BibitemOpen
  \bibfield  {author} {\bibinfo {author} {\bibfnamefont {G.~W.}\ \bibnamefont
  {Semenoff}}\ and\ \bibinfo {author} {\bibfnamefont {K.}~\bibnamefont
  {Zarembo}},\ }\href {\doibase 10.1103/PhysRevLett.107.171601} {\bibfield
  {journal} {\bibinfo  {journal} {Phys. Rev. Lett.}\ }\textbf {\bibinfo
  {volume} {107}},\ \bibinfo {pages} {171601} (\bibinfo {year} {2011})},\
  \Eprint {http://arxiv.org/abs/1109.2920} {arXiv:1109.2920 [hep-th]}
  \BibitemShut {NoStop}%
\bibitem [{\citenamefont {Li}\ \emph {et~al.}(2022)\citenamefont {Li},
  \citenamefont {Luo},\ and\ \citenamefont {Li}}]{Li:2022hka}%
  \BibitemOpen
  \bibfield  {author} {\bibinfo {author} {\bibfnamefont {S.-w.}\ \bibnamefont
  {Li}}, \bibinfo {author} {\bibfnamefont {S.-k.}\ \bibnamefont {Luo}}, \ and\
  \bibinfo {author} {\bibfnamefont {H.-q.}\ \bibnamefont {Li}},\ }\href
  {\doibase 10.1007/JHEP08(2022)206} {\bibfield  {journal} {\bibinfo  {journal}
  {JHEP}\ }\textbf {\bibinfo {volume} {08}},\ \bibinfo {pages} {206} (\bibinfo
  {year} {2022})},\ \Eprint {http://arxiv.org/abs/2205.01885} {arXiv:2205.01885
  [hep-th]} \BibitemShut {NoStop}%
\bibitem [{\citenamefont {Ghoroku}\ and\ \citenamefont
  {Ishihara}(2016)}]{Ghoroku:2016kft}%
  \BibitemOpen
  \bibfield  {author} {\bibinfo {author} {\bibfnamefont {K.}~\bibnamefont
  {Ghoroku}}\ and\ \bibinfo {author} {\bibfnamefont {M.}~\bibnamefont
  {Ishihara}},\ }\href {\doibase 10.1007/JHEP09(2016)011} {\bibfield  {journal}
  {\bibinfo  {journal} {JHEP}\ }\textbf {\bibinfo {volume} {09}},\ \bibinfo
  {pages} {011} (\bibinfo {year} {2016})},\ \Eprint
  {http://arxiv.org/abs/1604.05025} {arXiv:1604.05025 [hep-th]} \BibitemShut
  {NoStop}%
\bibitem [{\citenamefont {Zhang}\ \emph {et~al.}(2020)\citenamefont {Zhang},
  \citenamefont {Zhu},\ and\ \citenamefont {Hou}}]{Zhang:2020noe}%
  \BibitemOpen
  \bibfield  {author} {\bibinfo {author} {\bibfnamefont {Z.-Q.}\ \bibnamefont
  {Zhang}}, \bibinfo {author} {\bibfnamefont {X.}~\bibnamefont {Zhu}}, \ and\
  \bibinfo {author} {\bibfnamefont {D.-F.}\ \bibnamefont {Hou}},\ }\href
  {\doibase 10.1103/PhysRevD.101.026017} {\bibfield  {journal} {\bibinfo
  {journal} {Phys. Rev. D}\ }\textbf {\bibinfo {volume} {101}},\ \bibinfo
  {pages} {026017} (\bibinfo {year} {2020})},\ \Eprint
  {http://arxiv.org/abs/2001.02321} {arXiv:2001.02321 [hep-th]} \BibitemShut
  {NoStop}%
\bibitem [{\citenamefont {Cai}\ \emph {et~al.}(2022{\natexlab{a}})\citenamefont
  {Cai}, \citenamefont {Jing},\ and\ \citenamefont {Zhang}}]{Cai:2022nwq}%
  \BibitemOpen
  \bibfield  {author} {\bibinfo {author} {\bibfnamefont {Y.-z.}\ \bibnamefont
  {Cai}}, \bibinfo {author} {\bibfnamefont {R.-p.}\ \bibnamefont {Jing}}, \
  and\ \bibinfo {author} {\bibfnamefont {Z.-q.}\ \bibnamefont {Zhang}},\ }\href
  {\doibase 10.1088/1674-1137/ac7cd9} {\bibfield  {journal} {\bibinfo
  {journal} {Chin. Phys. C}\ }\textbf {\bibinfo {volume} {46}},\ \bibinfo
  {pages} {104107} (\bibinfo {year} {2022}{\natexlab{a}})},\ \Eprint
  {http://arxiv.org/abs/2206.15052} {arXiv:2206.15052 [hep-th]} \BibitemShut
  {NoStop}%
\bibitem [{\citenamefont {Cai}\ and\ \citenamefont
  {Zhang}(2024)}]{Cai:2023cjl}%
  \BibitemOpen
  \bibfield  {author} {\bibinfo {author} {\bibfnamefont {Y.-z.}\ \bibnamefont
  {Cai}}\ and\ \bibinfo {author} {\bibfnamefont {Z.-q.}\ \bibnamefont
  {Zhang}},\ }\href {\doibase 10.1088/1674-1137/ad061f} {\bibfield  {journal}
  {\bibinfo  {journal} {Chin. Phys. C}\ }\textbf {\bibinfo {volume} {48}},\
  \bibinfo {pages} {015102} (\bibinfo {year} {2024})},\ \Eprint
  {http://arxiv.org/abs/2310.13865} {arXiv:2310.13865 [hep-ph]} \BibitemShut
  {NoStop}%
\bibitem [{\citenamefont {Zhou}\ \emph {et~al.}(2024)\citenamefont {Zhou},
  \citenamefont {Chen}, \citenamefont {Zhang}, \citenamefont {Ping},\ and\
  \citenamefont {Chen}}]{Zhou:2021nbp}%
  \BibitemOpen
  \bibfield  {author} {\bibinfo {author} {\bibfnamefont {J.}~\bibnamefont
  {Zhou}}, \bibinfo {author} {\bibfnamefont {J.}~\bibnamefont {Chen}}, \bibinfo
  {author} {\bibfnamefont {L.}~\bibnamefont {Zhang}}, \bibinfo {author}
  {\bibfnamefont {J.}~\bibnamefont {Ping}}, \ and\ \bibinfo {author}
  {\bibfnamefont {X.}~\bibnamefont {Chen}},\ }\href {\doibase
  10.1140/epjc/s10052-024-12448-8} {\bibfield  {journal} {\bibinfo  {journal}
  {Eur. Phys. J. C}\ }\textbf {\bibinfo {volume} {84}},\ \bibinfo {pages} {94}
  (\bibinfo {year} {2024})},\ \Eprint {http://arxiv.org/abs/2101.08105}
  {arXiv:2101.08105 [hep-th]} \BibitemShut {NoStop}%
\bibitem [{\citenamefont {Zhu}\ \emph {et~al.}(2021)\citenamefont {Zhu},
  \citenamefont {Liu},\ and\ \citenamefont {Hou}}]{Zhu:2021ucv}%
  \BibitemOpen
  \bibfield  {author} {\bibinfo {author} {\bibfnamefont {Z.-R.}\ \bibnamefont
  {Zhu}}, \bibinfo {author} {\bibfnamefont {Y.-K.}\ \bibnamefont {Liu}}, \ and\
  \bibinfo {author} {\bibfnamefont {D.}~\bibnamefont {Hou}},\ }\href@noop {} {\
   (\bibinfo {year} {2021})},\ \Eprint {http://arxiv.org/abs/2108.05148}
  {arXiv:2108.05148 [hep-ph]} \BibitemShut {NoStop}%
\bibitem [{\citenamefont {Ding}\ and\ \citenamefont
  {Zhang}(2021)}]{Ding:2020slu}%
  \BibitemOpen
  \bibfield  {author} {\bibinfo {author} {\bibfnamefont {Y.}~\bibnamefont
  {Ding}}\ and\ \bibinfo {author} {\bibfnamefont {Z.-q.}\ \bibnamefont
  {Zhang}},\ }\href {\doibase 10.1088/1674-1137/abc240} {\bibfield  {journal}
  {\bibinfo  {journal} {Chin. Phys. C}\ }\textbf {\bibinfo {volume} {45}},\
  \bibinfo {pages} {013111} (\bibinfo {year} {2021})},\ \Eprint
  {http://arxiv.org/abs/2009.06179} {arXiv:2009.06179 [hep-th]} \BibitemShut
  {NoStop}%
\bibitem [{\citenamefont {Li}(2021)}]{Li:2020azb}%
  \BibitemOpen
  \bibfield  {author} {\bibinfo {author} {\bibfnamefont {S.-w.}\ \bibnamefont
  {Li}},\ }\href {\doibase 10.1140/epjc/s10052-021-09607-6} {\bibfield
  {journal} {\bibinfo  {journal} {Eur. Phys. J. C}\ }\textbf {\bibinfo {volume}
  {81}},\ \bibinfo {pages} {797} (\bibinfo {year} {2021})},\ \Eprint
  {http://arxiv.org/abs/2005.11955} {arXiv:2005.11955 [hep-th]} \BibitemShut
  {NoStop}%
\bibitem [{\citenamefont {Zhang}\ \emph {et~al.}(2017)\citenamefont {Zhang},
  \citenamefont {Hou},\ and\ \citenamefont {Chen}}]{Zhang:2017egb}%
  \BibitemOpen
  \bibfield  {author} {\bibinfo {author} {\bibfnamefont {Z.-q.}\ \bibnamefont
  {Zhang}}, \bibinfo {author} {\bibfnamefont {D.-f.}\ \bibnamefont {Hou}}, \
  and\ \bibinfo {author} {\bibfnamefont {G.}~\bibnamefont {Chen}},\ }\href
  {\doibase 10.1140/epja/i2017-12244-3} {\bibfield  {journal} {\bibinfo
  {journal} {Eur. Phys. J. A}\ }\textbf {\bibinfo {volume} {53}},\ \bibinfo
  {pages} {51} (\bibinfo {year} {2017})},\ \Eprint
  {http://arxiv.org/abs/1703.10213} {arXiv:1703.10213 [hep-th]} \BibitemShut
  {NoStop}%
\bibitem [{\citenamefont {Xu}\ \emph {et~al.}(2023)\citenamefont {Xu},
  \citenamefont {Ilyas},\ and\ \citenamefont {Huang}}]{Xu:2016cdt}%
  \BibitemOpen
  \bibfield  {author} {\bibinfo {author} {\bibfnamefont {H.}~\bibnamefont
  {Xu}}, \bibinfo {author} {\bibfnamefont {M.}~\bibnamefont {Ilyas}}, \ and\
  \bibinfo {author} {\bibfnamefont {Y.-C.}\ \bibnamefont {Huang}},\ }\href
  {\doibase 10.1155/2023/6614276} {\bibfield  {journal} {\bibinfo  {journal}
  {Adv. High Energy Phys.}\ }\textbf {\bibinfo {volume} {2023}},\ \bibinfo
  {pages} {6614276} (\bibinfo {year} {2023})},\ \Eprint
  {http://arxiv.org/abs/1604.06331} {arXiv:1604.06331 [hep-th]} \BibitemShut
  {NoStop}%
\bibitem [{\citenamefont {Zhang}\ \emph {et~al.}(2016)\citenamefont {Zhang},
  \citenamefont {Hou}, \citenamefont {Wu},\ and\ \citenamefont
  {Chen}}]{Zhang:2016qiz}%
  \BibitemOpen
  \bibfield  {author} {\bibinfo {author} {\bibfnamefont {Z.-q.}\ \bibnamefont
  {Zhang}}, \bibinfo {author} {\bibfnamefont {D.-f.}\ \bibnamefont {Hou}},
  \bibinfo {author} {\bibfnamefont {Y.}~\bibnamefont {Wu}}, \ and\ \bibinfo
  {author} {\bibfnamefont {G.}~\bibnamefont {Chen}},\ }\href {\doibase
  10.1155/2016/9258106} {\bibfield  {journal} {\bibinfo  {journal} {Adv. High
  Energy Phys.}\ }\textbf {\bibinfo {volume} {2016}},\ \bibinfo {pages}
  {9258106} (\bibinfo {year} {2016})},\ \Eprint
  {http://arxiv.org/abs/1604.00095} {arXiv:1604.00095 [hep-ph]} \BibitemShut
  {NoStop}%
\bibitem [{\citenamefont {Sadeghi}\ \emph {et~al.}(2017)\citenamefont
  {Sadeghi}, \citenamefont {Pourhassan}, \citenamefont {Tahery},\ and\
  \citenamefont {Razavi}}]{Sadeghi:2016ppy}%
  \BibitemOpen
  \bibfield  {author} {\bibinfo {author} {\bibfnamefont {J.}~\bibnamefont
  {Sadeghi}}, \bibinfo {author} {\bibfnamefont {B.}~\bibnamefont {Pourhassan}},
  \bibinfo {author} {\bibfnamefont {S.}~\bibnamefont {Tahery}}, \ and\ \bibinfo
  {author} {\bibfnamefont {F.}~\bibnamefont {Razavi}},\ }\href {\doibase
  10.1142/S0217751X17500452} {\bibfield  {journal} {\bibinfo  {journal} {Int.
  J. Mod. Phys. A}\ }\textbf {\bibinfo {volume} {32}},\ \bibinfo {pages}
  {1750045} (\bibinfo {year} {2017})},\ \Eprint
  {http://arxiv.org/abs/1603.07629} {arXiv:1603.07629 [hep-th]} \BibitemShut
  {NoStop}%
\bibitem [{\citenamefont {Shahkarami}\ \emph {et~al.}(2018)\citenamefont
  {Shahkarami}, \citenamefont {Dehghani},\ and\ \citenamefont
  {Dehghani}}]{Shahkarami:2015qff}%
  \BibitemOpen
  \bibfield  {author} {\bibinfo {author} {\bibfnamefont {L.}~\bibnamefont
  {Shahkarami}}, \bibinfo {author} {\bibfnamefont {M.}~\bibnamefont
  {Dehghani}}, \ and\ \bibinfo {author} {\bibfnamefont {P.}~\bibnamefont
  {Dehghani}},\ }\href {\doibase 10.1103/PhysRevD.97.046013} {\bibfield
  {journal} {\bibinfo  {journal} {Phys. Rev. D}\ }\textbf {\bibinfo {volume}
  {97}},\ \bibinfo {pages} {046013} (\bibinfo {year} {2018})},\ \Eprint
  {http://arxiv.org/abs/1511.07986} {arXiv:1511.07986 [hep-th]} \BibitemShut
  {NoStop}%
\bibitem [{\citenamefont {Zhang}\ and\ \citenamefont
  {Abdalla}(2016)}]{Zhang:2015bha}%
  \BibitemOpen
  \bibfield  {author} {\bibinfo {author} {\bibfnamefont {S.-J.}\ \bibnamefont
  {Zhang}}\ and\ \bibinfo {author} {\bibfnamefont {E.}~\bibnamefont
  {Abdalla}},\ }\href {\doibase 10.1007/s10714-016-2056-z} {\bibfield
  {journal} {\bibinfo  {journal} {Gen. Rel. Grav.}\ }\textbf {\bibinfo {volume}
  {48}},\ \bibinfo {pages} {60} (\bibinfo {year} {2016})},\ \Eprint
  {http://arxiv.org/abs/1508.03364} {arXiv:1508.03364 [hep-th]} \BibitemShut
  {NoStop}%
\bibitem [{\citenamefont {Kawai}\ \emph {et~al.}(2015)\citenamefont {Kawai},
  \citenamefont {Sato},\ and\ \citenamefont {Yoshida}}]{Kawai:2015mha}%
  \BibitemOpen
  \bibfield  {author} {\bibinfo {author} {\bibfnamefont {D.}~\bibnamefont
  {Kawai}}, \bibinfo {author} {\bibfnamefont {Y.}~\bibnamefont {Sato}}, \ and\
  \bibinfo {author} {\bibfnamefont {K.}~\bibnamefont {Yoshida}},\ }\href
  {\doibase 10.1142/S0217751X15300264} {\bibfield  {journal} {\bibinfo
  {journal} {Int. J. Mod. Phys. A}\ }\textbf {\bibinfo {volume} {30}},\
  \bibinfo {pages} {1530026} (\bibinfo {year} {2015})},\ \Eprint
  {http://arxiv.org/abs/1504.00459} {arXiv:1504.00459 [hep-th]} \BibitemShut
  {NoStop}%
\bibitem [{\citenamefont {Chakrabortty}\ and\ \citenamefont
  {Sathiapalan}(2014)}]{Chakrabortty:2014kma}%
  \BibitemOpen
  \bibfield  {author} {\bibinfo {author} {\bibfnamefont {S.}~\bibnamefont
  {Chakrabortty}}\ and\ \bibinfo {author} {\bibfnamefont {B.}~\bibnamefont
  {Sathiapalan}},\ }\href {\doibase 10.1016/j.nuclphysb.2014.11.010} {\bibfield
   {journal} {\bibinfo  {journal} {Nucl. Phys. B}\ }\textbf {\bibinfo {volume}
  {890}},\ \bibinfo {pages} {241} (\bibinfo {year} {2014})},\ \Eprint
  {http://arxiv.org/abs/1409.1383} {arXiv:1409.1383 [hep-th]} \BibitemShut
  {NoStop}%
\bibitem [{\citenamefont {Sato}\ and\ \citenamefont
  {Yoshida}(2013{\natexlab{a}})}]{Sato:2013hyw}%
  \BibitemOpen
  \bibfield  {author} {\bibinfo {author} {\bibfnamefont {Y.}~\bibnamefont
  {Sato}}\ and\ \bibinfo {author} {\bibfnamefont {K.}~\bibnamefont {Yoshida}},\
  }\href {\doibase 10.1007/JHEP12(2013)051} {\bibfield  {journal} {\bibinfo
  {journal} {JHEP}\ }\textbf {\bibinfo {volume} {12}},\ \bibinfo {pages} {051}
  (\bibinfo {year} {2013}{\natexlab{a}})},\ \Eprint
  {http://arxiv.org/abs/1309.4629} {arXiv:1309.4629 [hep-th]} \BibitemShut
  {NoStop}%
\bibitem [{\citenamefont {Sato}\ and\ \citenamefont
  {Yoshida}(2013{\natexlab{b}})}]{Sato:2013dwa}%
  \BibitemOpen
  \bibfield  {author} {\bibinfo {author} {\bibfnamefont {Y.}~\bibnamefont
  {Sato}}\ and\ \bibinfo {author} {\bibfnamefont {K.}~\bibnamefont {Yoshida}},\
  }\href {\doibase 10.1007/JHEP09(2013)134} {\bibfield  {journal} {\bibinfo
  {journal} {JHEP}\ }\textbf {\bibinfo {volume} {09}},\ \bibinfo {pages} {134}
  (\bibinfo {year} {2013}{\natexlab{b}})},\ \Eprint
  {http://arxiv.org/abs/1306.5512} {arXiv:1306.5512 [hep-th]} \BibitemShut
  {NoStop}%
\bibitem [{\citenamefont {Fischler}\ \emph {et~al.}(2015)\citenamefont
  {Fischler}, \citenamefont {Nguyen}, \citenamefont {Pedraza},\ and\
  \citenamefont {Tangarife}}]{Fischler:2014ama}%
  \BibitemOpen
  \bibfield  {author} {\bibinfo {author} {\bibfnamefont {W.}~\bibnamefont
  {Fischler}}, \bibinfo {author} {\bibfnamefont {P.~H.}\ \bibnamefont
  {Nguyen}}, \bibinfo {author} {\bibfnamefont {J.~F.}\ \bibnamefont {Pedraza}},
  \ and\ \bibinfo {author} {\bibfnamefont {W.}~\bibnamefont {Tangarife}},\
  }\href {\doibase 10.1103/PhysRevD.91.086015} {\bibfield  {journal} {\bibinfo
  {journal} {Phys. Rev. D}\ }\textbf {\bibinfo {volume} {91}},\ \bibinfo
  {pages} {086015} (\bibinfo {year} {2015})},\ \Eprint
  {http://arxiv.org/abs/1411.1787} {arXiv:1411.1787 [hep-th]} \BibitemShut
  {NoStop}%
\bibitem [{\citenamefont {Bitaghsir~Fadafan}\ and\ \citenamefont
  {Saiedi}(2015)}]{BitaghsirFadafan:2015asm}%
  \BibitemOpen
  \bibfield  {author} {\bibinfo {author} {\bibfnamefont {K.}~\bibnamefont
  {Bitaghsir~Fadafan}}\ and\ \bibinfo {author} {\bibfnamefont {F.}~\bibnamefont
  {Saiedi}},\ }\href {\doibase 10.1140/epjc/s10052-015-3839-1} {\bibfield
  {journal} {\bibinfo  {journal} {Eur. Phys. J. C}\ }\textbf {\bibinfo {volume}
  {75}},\ \bibinfo {pages} {612} (\bibinfo {year} {2015})},\ \Eprint
  {http://arxiv.org/abs/1504.02432} {arXiv:1504.02432 [hep-th]} \BibitemShut
  {NoStop}%
\bibitem [{\citenamefont {Zhu}\ \emph {et~al.}(2024)\citenamefont {Zhu},
  \citenamefont {Sun},\ and\ \citenamefont {Han}}]{Zhu:2024pdx}%
  \BibitemOpen
  \bibfield  {author} {\bibinfo {author} {\bibfnamefont {Z.-R.}\ \bibnamefont
  {Zhu}}, \bibinfo {author} {\bibfnamefont {M.}~\bibnamefont {Sun}}, \ and\
  \bibinfo {author} {\bibfnamefont {J.}~\bibnamefont {Han}},\ }\href {\doibase
  10.1140/epjp/s13360-024-05253-5} {\bibfield  {journal} {\bibinfo  {journal}
  {Eur. Phys. J. Plus}\ }\textbf {\bibinfo {volume} {139}},\ \bibinfo {pages}
  {420} (\bibinfo {year} {2024})}\BibitemShut {NoStop}%
\bibitem [{\citenamefont {Wu}(2015)}]{Wu:2015krf}%
  \BibitemOpen
  \bibfield  {author} {\bibinfo {author} {\bibfnamefont {X.}~\bibnamefont
  {Wu}},\ }\href {\doibase 10.1007/JHEP09(2015)044} {\bibfield  {journal}
  {\bibinfo  {journal} {JHEP}\ }\textbf {\bibinfo {volume} {09}},\ \bibinfo
  {pages} {044} (\bibinfo {year} {2015})},\ \Eprint
  {http://arxiv.org/abs/1507.03208} {arXiv:1507.03208 [hep-th]} \BibitemShut
  {NoStop}%
\bibitem [{\citenamefont {Sato}\ and\ \citenamefont
  {Yoshida}(2013{\natexlab{c}})}]{Sato:2013pxa}%
  \BibitemOpen
  \bibfield  {author} {\bibinfo {author} {\bibfnamefont {Y.}~\bibnamefont
  {Sato}}\ and\ \bibinfo {author} {\bibfnamefont {K.}~\bibnamefont {Yoshida}},\
  }\href {\doibase 10.1007/JHEP04(2013)111} {\bibfield  {journal} {\bibinfo
  {journal} {JHEP}\ }\textbf {\bibinfo {volume} {04}},\ \bibinfo {pages} {111}
  (\bibinfo {year} {2013}{\natexlab{c}})},\ \Eprint
  {http://arxiv.org/abs/1303.0112} {arXiv:1303.0112 [hep-th]} \BibitemShut
  {NoStop}%
\bibitem [{\citenamefont {Borsanyi}\ \emph {et~al.}(2010)\citenamefont
  {Borsanyi}, \citenamefont {Endrodi}, \citenamefont {Fodor}, \citenamefont
  {Jakovac}, \citenamefont {Katz}, \citenamefont {Krieg}, \citenamefont
  {Ratti},\ and\ \citenamefont {Szabo}}]{Borsanyi:2010cj}%
  \BibitemOpen
  \bibfield  {author} {\bibinfo {author} {\bibfnamefont {S.}~\bibnamefont
  {Borsanyi}}, \bibinfo {author} {\bibfnamefont {G.}~\bibnamefont {Endrodi}},
  \bibinfo {author} {\bibfnamefont {Z.}~\bibnamefont {Fodor}}, \bibinfo
  {author} {\bibfnamefont {A.}~\bibnamefont {Jakovac}}, \bibinfo {author}
  {\bibfnamefont {S.~D.}\ \bibnamefont {Katz}}, \bibinfo {author}
  {\bibfnamefont {S.}~\bibnamefont {Krieg}}, \bibinfo {author} {\bibfnamefont
  {C.}~\bibnamefont {Ratti}}, \ and\ \bibinfo {author} {\bibfnamefont {K.~K.}\
  \bibnamefont {Szabo}},\ }\href {\doibase 10.1007/JHEP11(2010)077} {\bibfield
  {journal} {\bibinfo  {journal} {JHEP}\ }\textbf {\bibinfo {volume} {11}},\
  \bibinfo {pages} {077} (\bibinfo {year} {2010})},\ \Eprint
  {http://arxiv.org/abs/1007.2580} {arXiv:1007.2580 [hep-lat]} \BibitemShut
  {NoStop}%
\bibitem [{\citenamefont {Borsanyi}\ \emph {et~al.}(2011)\citenamefont
  {Borsanyi}, \citenamefont {Endrodi}, \citenamefont {Fodor}, \citenamefont
  {Katz}, \citenamefont {Krieg}, \citenamefont {Ratti}, \citenamefont
  {Schroeder},\ and\ \citenamefont {Szabo}}]{Borsanyi:2011wyg}%
  \BibitemOpen
  \bibfield  {author} {\bibinfo {author} {\bibfnamefont {S.}~\bibnamefont
  {Borsanyi}}, \bibinfo {author} {\bibfnamefont {G.}~\bibnamefont {Endrodi}},
  \bibinfo {author} {\bibfnamefont {Z.}~\bibnamefont {Fodor}}, \bibinfo
  {author} {\bibfnamefont {S.~D.}\ \bibnamefont {Katz}}, \bibinfo {author}
  {\bibfnamefont {S.}~\bibnamefont {Krieg}}, \bibinfo {author} {\bibfnamefont
  {C.}~\bibnamefont {Ratti}}, \bibinfo {author} {\bibfnamefont
  {C.}~\bibnamefont {Schroeder}}, \ and\ \bibinfo {author} {\bibfnamefont
  {K.~K.}\ \bibnamefont {Szabo}},\ }\href {\doibase 10.22323/1.139.0201}
  {\bibfield  {journal} {\bibinfo  {journal} {PoS}\ }\textbf {\bibinfo {volume}
  {LATTICE2011}},\ \bibinfo {pages} {201} (\bibinfo {year} {2011})},\ \Eprint
  {http://arxiv.org/abs/1204.0995} {arXiv:1204.0995 [hep-lat]} \BibitemShut
  {NoStop}%
\bibitem [{\citenamefont {Borsanyi}\ \emph {et~al.}(2012)\citenamefont
  {Borsanyi}, \citenamefont {Endrodi}, \citenamefont {Fodor}, \citenamefont
  {Katz},\ and\ \citenamefont {Szabo}}]{Borsanyi:2012ve}%
  \BibitemOpen
  \bibfield  {author} {\bibinfo {author} {\bibfnamefont {S.}~\bibnamefont
  {Borsanyi}}, \bibinfo {author} {\bibfnamefont {G.}~\bibnamefont {Endrodi}},
  \bibinfo {author} {\bibfnamefont {Z.}~\bibnamefont {Fodor}}, \bibinfo
  {author} {\bibfnamefont {S.~D.}\ \bibnamefont {Katz}}, \ and\ \bibinfo
  {author} {\bibfnamefont {K.~K.}\ \bibnamefont {Szabo}},\ }\href {\doibase
  10.1007/JHEP07(2012)056} {\bibfield  {journal} {\bibinfo  {journal} {JHEP}\
  }\textbf {\bibinfo {volume} {07}},\ \bibinfo {pages} {056} (\bibinfo {year}
  {2012})},\ \Eprint {http://arxiv.org/abs/1204.6184} {arXiv:1204.6184
  [hep-lat]} \BibitemShut {NoStop}%
\bibitem [{\citenamefont {Fischer}(2019)}]{Fischer:2018sdj}%
  \BibitemOpen
  \bibfield  {author} {\bibinfo {author} {\bibfnamefont {C.~S.}\ \bibnamefont
  {Fischer}},\ }\href {\doibase 10.1016/j.ppnp.2019.01.002} {\bibfield
  {journal} {\bibinfo  {journal} {Prog. Part. Nucl. Phys.}\ }\textbf {\bibinfo
  {volume} {105}},\ \bibinfo {pages} {1} (\bibinfo {year} {2019})},\ \Eprint
  {http://arxiv.org/abs/1810.12938} {arXiv:1810.12938 [hep-ph]} \BibitemShut
  {NoStop}%
\bibitem [{\citenamefont {Ratti}\ \emph {et~al.}(2013)\citenamefont {Ratti},
  \citenamefont {Borsanyi}, \citenamefont {Endrodi}, \citenamefont {Fodor},
  \citenamefont {Katz}, \citenamefont {Krieg}, \citenamefont {Schroeder},\ and\
  \citenamefont {Szabo}}]{Ratti:2013uta}%
  \BibitemOpen
  \bibfield  {author} {\bibinfo {author} {\bibfnamefont {C.}~\bibnamefont
  {Ratti}}, \bibinfo {author} {\bibfnamefont {S.}~\bibnamefont {Borsanyi}},
  \bibinfo {author} {\bibfnamefont {G.}~\bibnamefont {Endrodi}}, \bibinfo
  {author} {\bibfnamefont {Z.}~\bibnamefont {Fodor}}, \bibinfo {author}
  {\bibfnamefont {S.~D.}\ \bibnamefont {Katz}}, \bibinfo {author}
  {\bibfnamefont {S.}~\bibnamefont {Krieg}}, \bibinfo {author} {\bibfnamefont
  {C.}~\bibnamefont {Schroeder}}, \ and\ \bibinfo {author} {\bibfnamefont
  {K.~K.}\ \bibnamefont {Szabo}},\ }\href {\doibase
  10.1016/j.nuclphysa.2013.02.153} {\bibfield  {journal} {\bibinfo  {journal}
  {Nucl. Phys. A}\ }\textbf {\bibinfo {volume} {904-905}},\ \bibinfo {pages}
  {869c} (\bibinfo {year} {2013})}\BibitemShut {NoStop}%
\bibitem [{\citenamefont {Weber}\ \emph {et~al.}(2021)\citenamefont {Weber},
  \citenamefont {Bazavov},\ and\ \citenamefont {Petreczky}}]{Weber:2021hro}%
  \BibitemOpen
  \bibfield  {author} {\bibinfo {author} {\bibfnamefont {J.~H.}\ \bibnamefont
  {Weber}}, \bibinfo {author} {\bibfnamefont {A.}~\bibnamefont {Bazavov}}, \
  and\ \bibinfo {author} {\bibfnamefont {P.}~\bibnamefont {Petreczky}},\ }\href
  {\doibase 10.22323/1.396.0060} {\bibfield  {journal} {\bibinfo  {journal}
  {PoS}\ }\textbf {\bibinfo {volume} {LATTICE2021}},\ \bibinfo {pages} {060}
  (\bibinfo {year} {2021})},\ \Eprint {http://arxiv.org/abs/2110.03606}
  {arXiv:2110.03606 [hep-lat]} \BibitemShut {NoStop}%
\bibitem [{\citenamefont {Datta}\ \emph {et~al.}(2017)\citenamefont {Datta},
  \citenamefont {Gavai},\ and\ \citenamefont {Gupta}}]{Datta:2016ukp}%
  \BibitemOpen
  \bibfield  {author} {\bibinfo {author} {\bibfnamefont {S.}~\bibnamefont
  {Datta}}, \bibinfo {author} {\bibfnamefont {R.~V.}\ \bibnamefont {Gavai}}, \
  and\ \bibinfo {author} {\bibfnamefont {S.}~\bibnamefont {Gupta}},\ }\href
  {\doibase 10.1103/PhysRevD.95.054512} {\bibfield  {journal} {\bibinfo
  {journal} {Phys. Rev. D}\ }\textbf {\bibinfo {volume} {95}},\ \bibinfo
  {pages} {054512} (\bibinfo {year} {2017})},\ \Eprint
  {http://arxiv.org/abs/1612.06673} {arXiv:1612.06673 [hep-lat]} \BibitemShut
  {NoStop}%
\bibitem [{\citenamefont {Bellwied}\ \emph {et~al.}(2015)\citenamefont
  {Bellwied}, \citenamefont {Borsanyi}, \citenamefont {Fodor}, \citenamefont
  {Katz}, \citenamefont {Pasztor}, \citenamefont {Ratti},\ and\ \citenamefont
  {Szabo}}]{Bellwied:2015lba}%
  \BibitemOpen
  \bibfield  {author} {\bibinfo {author} {\bibfnamefont {R.}~\bibnamefont
  {Bellwied}}, \bibinfo {author} {\bibfnamefont {S.}~\bibnamefont {Borsanyi}},
  \bibinfo {author} {\bibfnamefont {Z.}~\bibnamefont {Fodor}}, \bibinfo
  {author} {\bibfnamefont {S.~D.}\ \bibnamefont {Katz}}, \bibinfo {author}
  {\bibfnamefont {A.}~\bibnamefont {Pasztor}}, \bibinfo {author} {\bibfnamefont
  {C.}~\bibnamefont {Ratti}}, \ and\ \bibinfo {author} {\bibfnamefont {K.~K.}\
  \bibnamefont {Szabo}},\ }\href {\doibase 10.1103/PhysRevD.92.114505}
  {\bibfield  {journal} {\bibinfo  {journal} {Phys. Rev. D}\ }\textbf {\bibinfo
  {volume} {92}},\ \bibinfo {pages} {114505} (\bibinfo {year} {2015})},\
  \Eprint {http://arxiv.org/abs/1507.04627} {arXiv:1507.04627 [hep-lat]}
  \BibitemShut {NoStop}%
\bibitem [{\citenamefont {Bazavov}\ \emph {et~al.}(2017)\citenamefont {Bazavov}
  \emph {et~al.}}]{Bazavov:2017dus}%
  \BibitemOpen
  \bibfield  {author} {\bibinfo {author} {\bibfnamefont {A.}~\bibnamefont
  {Bazavov}} \emph {et~al.},\ }\href {\doibase 10.1103/PhysRevD.95.054504}
  {\bibfield  {journal} {\bibinfo  {journal} {Phys. Rev. D}\ }\textbf {\bibinfo
  {volume} {95}},\ \bibinfo {pages} {054504} (\bibinfo {year} {2017})},\
  \Eprint {http://arxiv.org/abs/1701.04325} {arXiv:1701.04325 [hep-lat]}
  \BibitemShut {NoStop}%
\bibitem [{\citenamefont {Sharma}(2023)}]{Sharma:2022ztl}%
  \BibitemOpen
  \bibfield  {author} {\bibinfo {author} {\bibfnamefont {S.}~\bibnamefont
  {Sharma}},\ }\href {\doibase 10.22323/1.430.0191} {\bibfield  {journal}
  {\bibinfo  {journal} {PoS}\ }\textbf {\bibinfo {volume} {LATTICE2022}},\
  \bibinfo {pages} {191} (\bibinfo {year} {2023})},\ \Eprint
  {http://arxiv.org/abs/2212.11148} {arXiv:2212.11148 [hep-lat]} \BibitemShut
  {NoStop}%
\bibitem [{\citenamefont {Chen}\ and\ \citenamefont
  {Huang}(2024{\natexlab{a}})}]{Chen:2024ckb}%
  \BibitemOpen
  \bibfield  {author} {\bibinfo {author} {\bibfnamefont {X.}~\bibnamefont
  {Chen}}\ and\ \bibinfo {author} {\bibfnamefont {M.}~\bibnamefont {Huang}},\
  }\href {\doibase 10.1103/PhysRevD.109.L051902} {\bibfield  {journal}
  {\bibinfo  {journal} {Phys. Rev. D}\ }\textbf {\bibinfo {volume} {109}},\
  \bibinfo {pages} {L051902} (\bibinfo {year} {2024}{\natexlab{a}})},\ \Eprint
  {http://arxiv.org/abs/2401.06417} {arXiv:2401.06417 [hep-ph]} \BibitemShut
  {NoStop}%
\bibitem [{\citenamefont {Chen}\ and\ \citenamefont
  {Huang}(2024{\natexlab{b}})}]{Chen:2024mmd}%
  \BibitemOpen
  \bibfield  {author} {\bibinfo {author} {\bibfnamefont {X.}~\bibnamefont
  {Chen}}\ and\ \bibinfo {author} {\bibfnamefont {M.}~\bibnamefont {Huang}},\
  }\href@noop {} {\  (\bibinfo {year} {2024}{\natexlab{b}})},\ \Eprint
  {http://arxiv.org/abs/2405.06179} {arXiv:2405.06179 [hep-ph]} \BibitemShut
  {NoStop}%
\bibitem [{\citenamefont {Guo}\ \emph {et~al.}(2024)\citenamefont {Guo},
  \citenamefont {Chen}, \citenamefont {Xiang}, \citenamefont
  {Martin~Contreras},\ and\ \citenamefont {Li}}]{Guo:2024qiq}%
  \BibitemOpen
  \bibfield  {author} {\bibinfo {author} {\bibfnamefont {X.}~\bibnamefont
  {Guo}}, \bibinfo {author} {\bibfnamefont {X.}~\bibnamefont {Chen}}, \bibinfo
  {author} {\bibfnamefont {D.}~\bibnamefont {Xiang}}, \bibinfo {author}
  {\bibfnamefont {M.~A.}\ \bibnamefont {Martin~Contreras}}, \ and\ \bibinfo
  {author} {\bibfnamefont {X.-H.}\ \bibnamefont {Li}},\ }\href@noop {} {\
  (\bibinfo {year} {2024})},\ \Eprint {http://arxiv.org/abs/2406.04650}
  {arXiv:2406.04650 [hep-ph]} \BibitemShut {NoStop}%
\bibitem [{\citenamefont {Chen}\ \emph {et~al.}(2024)\citenamefont {Chen},
  \citenamefont {Chen}, \citenamefont {Li}, \citenamefont {Zhu},\ and\
  \citenamefont {Zhou}}]{Chen:2024epd}%
  \BibitemOpen
  \bibfield  {author} {\bibinfo {author} {\bibfnamefont {B.}~\bibnamefont
  {Chen}}, \bibinfo {author} {\bibfnamefont {X.}~\bibnamefont {Chen}}, \bibinfo
  {author} {\bibfnamefont {X.}~\bibnamefont {Li}}, \bibinfo {author}
  {\bibfnamefont {Z.-R.}\ \bibnamefont {Zhu}}, \ and\ \bibinfo {author}
  {\bibfnamefont {K.}~\bibnamefont {Zhou}},\ }\href@noop {} {\  (\bibinfo
  {year} {2024})},\ \Eprint {http://arxiv.org/abs/2404.18217} {arXiv:2404.18217
  [hep-ph]} \BibitemShut {NoStop}%
\bibitem [{\citenamefont {DeWolfe}\ \emph
  {et~al.}(2011{\natexlab{a}})\citenamefont {DeWolfe}, \citenamefont {Gubser},\
  and\ \citenamefont {Rosen}}]{DeWolfe:2010he}%
  \BibitemOpen
  \bibfield  {author} {\bibinfo {author} {\bibfnamefont {O.}~\bibnamefont
  {DeWolfe}}, \bibinfo {author} {\bibfnamefont {S.~S.}\ \bibnamefont {Gubser}},
  \ and\ \bibinfo {author} {\bibfnamefont {C.}~\bibnamefont {Rosen}},\ }\href
  {\doibase 10.1103/PhysRevD.83.086005} {\bibfield  {journal} {\bibinfo
  {journal} {Phys. Rev. D}\ }\textbf {\bibinfo {volume} {83}},\ \bibinfo
  {pages} {086005} (\bibinfo {year} {2011}{\natexlab{a}})},\ \Eprint
  {http://arxiv.org/abs/1012.1864} {arXiv:1012.1864 [hep-th]} \BibitemShut
  {NoStop}%
\bibitem [{\citenamefont {DeWolfe}\ \emph
  {et~al.}(2011{\natexlab{b}})\citenamefont {DeWolfe}, \citenamefont {Gubser},\
  and\ \citenamefont {Rosen}}]{DeWolfe:2011ts}%
  \BibitemOpen
  \bibfield  {author} {\bibinfo {author} {\bibfnamefont {O.}~\bibnamefont
  {DeWolfe}}, \bibinfo {author} {\bibfnamefont {S.~S.}\ \bibnamefont {Gubser}},
  \ and\ \bibinfo {author} {\bibfnamefont {C.}~\bibnamefont {Rosen}},\ }\href
  {\doibase 10.1103/PhysRevD.84.126014} {\bibfield  {journal} {\bibinfo
  {journal} {Phys. Rev. D}\ }\textbf {\bibinfo {volume} {84}},\ \bibinfo
  {pages} {126014} (\bibinfo {year} {2011}{\natexlab{b}})},\ \Eprint
  {http://arxiv.org/abs/1108.2029} {arXiv:1108.2029 [hep-th]} \BibitemShut
  {NoStop}%
\bibitem [{\citenamefont {Rougemont}\ \emph {et~al.}(2016)\citenamefont
  {Rougemont}, \citenamefont {Ficnar}, \citenamefont {Finazzo},\ and\
  \citenamefont {Noronha}}]{Rougemont:2015wca}%
  \BibitemOpen
  \bibfield  {author} {\bibinfo {author} {\bibfnamefont {R.}~\bibnamefont
  {Rougemont}}, \bibinfo {author} {\bibfnamefont {A.}~\bibnamefont {Ficnar}},
  \bibinfo {author} {\bibfnamefont {S.}~\bibnamefont {Finazzo}}, \ and\
  \bibinfo {author} {\bibfnamefont {J.}~\bibnamefont {Noronha}},\ }\href
  {\doibase 10.1007/JHEP04(2016)102} {\bibfield  {journal} {\bibinfo  {journal}
  {JHEP}\ }\textbf {\bibinfo {volume} {04}},\ \bibinfo {pages} {102} (\bibinfo
  {year} {2016})},\ \Eprint {http://arxiv.org/abs/1507.06556} {arXiv:1507.06556
  [hep-th]} \BibitemShut {NoStop}%
\bibitem [{\citenamefont {Rougemont}\ \emph {et~al.}(2017)\citenamefont
  {Rougemont}, \citenamefont {Critelli}, \citenamefont {Noronha-Hostler},
  \citenamefont {Noronha},\ and\ \citenamefont {Ratti}}]{Rougemont:2017tlu}%
  \BibitemOpen
  \bibfield  {author} {\bibinfo {author} {\bibfnamefont {R.}~\bibnamefont
  {Rougemont}}, \bibinfo {author} {\bibfnamefont {R.}~\bibnamefont {Critelli}},
  \bibinfo {author} {\bibfnamefont {J.}~\bibnamefont {Noronha-Hostler}},
  \bibinfo {author} {\bibfnamefont {J.}~\bibnamefont {Noronha}}, \ and\
  \bibinfo {author} {\bibfnamefont {C.}~\bibnamefont {Ratti}},\ }\href
  {\doibase 10.1103/PhysRevD.96.014032} {\bibfield  {journal} {\bibinfo
  {journal} {Phys. Rev. D}\ }\textbf {\bibinfo {volume} {96}},\ \bibinfo
  {pages} {014032} (\bibinfo {year} {2017})},\ \Eprint
  {http://arxiv.org/abs/1704.05558} {arXiv:1704.05558 [hep-ph]} \BibitemShut
  {NoStop}%
\bibitem [{\citenamefont {Grefa}\ \emph {et~al.}(2022)\citenamefont {Grefa},
  \citenamefont {Hippert}, \citenamefont {Noronha}, \citenamefont
  {Noronha-Hostler}, \citenamefont {Portillo}, \citenamefont {Ratti},\ and\
  \citenamefont {Rougemont}}]{Grefa:2022sav}%
  \BibitemOpen
  \bibfield  {author} {\bibinfo {author} {\bibfnamefont {J.}~\bibnamefont
  {Grefa}}, \bibinfo {author} {\bibfnamefont {M.}~\bibnamefont {Hippert}},
  \bibinfo {author} {\bibfnamefont {J.}~\bibnamefont {Noronha}}, \bibinfo
  {author} {\bibfnamefont {J.}~\bibnamefont {Noronha-Hostler}}, \bibinfo
  {author} {\bibfnamefont {I.}~\bibnamefont {Portillo}}, \bibinfo {author}
  {\bibfnamefont {C.}~\bibnamefont {Ratti}}, \ and\ \bibinfo {author}
  {\bibfnamefont {R.}~\bibnamefont {Rougemont}},\ }\href {\doibase
  10.1103/PhysRevD.106.034024} {\bibfield  {journal} {\bibinfo  {journal}
  {Phys. Rev. D}\ }\textbf {\bibinfo {volume} {106}},\ \bibinfo {pages}
  {034024} (\bibinfo {year} {2022})},\ \Eprint
  {http://arxiv.org/abs/2203.00139} {arXiv:2203.00139 [nucl-th]} \BibitemShut
  {NoStop}%
\bibitem [{\citenamefont {Zhou}\ \emph {et~al.}(2020)\citenamefont {Zhou},
  \citenamefont {Chen}, \citenamefont {Zhao},\ and\ \citenamefont
  {Ping}}]{Zhou:2020ssi}%
  \BibitemOpen
  \bibfield  {author} {\bibinfo {author} {\bibfnamefont {J.}~\bibnamefont
  {Zhou}}, \bibinfo {author} {\bibfnamefont {X.}~\bibnamefont {Chen}}, \bibinfo
  {author} {\bibfnamefont {Y.-Q.}\ \bibnamefont {Zhao}}, \ and\ \bibinfo
  {author} {\bibfnamefont {J.}~\bibnamefont {Ping}},\ }\href {\doibase
  10.1103/PhysRevD.102.086020} {\bibfield  {journal} {\bibinfo  {journal}
  {Phys. Rev. D}\ }\textbf {\bibinfo {volume} {102}},\ \bibinfo {pages}
  {086020} (\bibinfo {year} {2020})},\ \Eprint
  {http://arxiv.org/abs/2006.09062} {arXiv:2006.09062 [hep-ph]} \BibitemShut
  {NoStop}%
\bibitem [{\citenamefont {He}\ \emph {et~al.}(2013)\citenamefont {He},
  \citenamefont {Wu}, \citenamefont {Yang},\ and\ \citenamefont
  {Yuan}}]{He:2013qq}%
  \BibitemOpen
  \bibfield  {author} {\bibinfo {author} {\bibfnamefont {S.}~\bibnamefont
  {He}}, \bibinfo {author} {\bibfnamefont {S.-Y.}\ \bibnamefont {Wu}}, \bibinfo
  {author} {\bibfnamefont {Y.}~\bibnamefont {Yang}}, \ and\ \bibinfo {author}
  {\bibfnamefont {P.-H.}\ \bibnamefont {Yuan}},\ }\href {\doibase
  10.1007/JHEP04(2013)093} {\bibfield  {journal} {\bibinfo  {journal} {JHEP}\
  }\textbf {\bibinfo {volume} {04}},\ \bibinfo {pages} {093} (\bibinfo {year}
  {2013})},\ \Eprint {http://arxiv.org/abs/1301.0385} {arXiv:1301.0385
  [hep-th]} \BibitemShut {NoStop}%
\bibitem [{\citenamefont {Yang}\ and\ \citenamefont
  {Yuan}(2014)}]{Yang:2014bqa}%
  \BibitemOpen
  \bibfield  {author} {\bibinfo {author} {\bibfnamefont {Y.}~\bibnamefont
  {Yang}}\ and\ \bibinfo {author} {\bibfnamefont {P.-H.}\ \bibnamefont
  {Yuan}},\ }\href {\doibase 10.1007/JHEP11(2014)149} {\bibfield  {journal}
  {\bibinfo  {journal} {JHEP}\ }\textbf {\bibinfo {volume} {11}},\ \bibinfo
  {pages} {149} (\bibinfo {year} {2014})},\ \Eprint
  {http://arxiv.org/abs/1406.1865} {arXiv:1406.1865 [hep-th]} \BibitemShut
  {NoStop}%
\bibitem [{\citenamefont {Yang}\ and\ \citenamefont
  {Yuan}(2015)}]{Yang:2015aia}%
  \BibitemOpen
  \bibfield  {author} {\bibinfo {author} {\bibfnamefont {Y.}~\bibnamefont
  {Yang}}\ and\ \bibinfo {author} {\bibfnamefont {P.-H.}\ \bibnamefont
  {Yuan}},\ }\href {\doibase 10.1007/JHEP12(2015)161} {\bibfield  {journal}
  {\bibinfo  {journal} {JHEP}\ }\textbf {\bibinfo {volume} {12}},\ \bibinfo
  {pages} {161} (\bibinfo {year} {2015})},\ \Eprint
  {http://arxiv.org/abs/1506.05930} {arXiv:1506.05930 [hep-th]} \BibitemShut
  {NoStop}%
\bibitem [{\citenamefont {Dudal}\ and\ \citenamefont
  {Mahapatra}(2018)}]{Dudal:2018ztm}%
  \BibitemOpen
  \bibfield  {author} {\bibinfo {author} {\bibfnamefont {D.}~\bibnamefont
  {Dudal}}\ and\ \bibinfo {author} {\bibfnamefont {S.}~\bibnamefont
  {Mahapatra}},\ }\href {\doibase 10.1007/JHEP07(2018)120} {\bibfield
  {journal} {\bibinfo  {journal} {JHEP}\ }\textbf {\bibinfo {volume} {07}},\
  \bibinfo {pages} {120} (\bibinfo {year} {2018})},\ \Eprint
  {http://arxiv.org/abs/1805.02938} {arXiv:1805.02938 [hep-th]} \BibitemShut
  {NoStop}%
\bibitem [{\citenamefont {Chen}\ \emph {et~al.}(2019)\citenamefont {Chen},
  \citenamefont {Li},\ and\ \citenamefont {Huang}}]{Chen:2018vty}%
  \BibitemOpen
  \bibfield  {author} {\bibinfo {author} {\bibfnamefont {X.}~\bibnamefont
  {Chen}}, \bibinfo {author} {\bibfnamefont {D.}~\bibnamefont {Li}}, \ and\
  \bibinfo {author} {\bibfnamefont {M.}~\bibnamefont {Huang}},\ }\href
  {\doibase 10.1088/1674-1137/43/2/023105} {\bibfield  {journal} {\bibinfo
  {journal} {Chin. Phys. C}\ }\textbf {\bibinfo {volume} {43}},\ \bibinfo
  {pages} {023105} (\bibinfo {year} {2019})},\ \Eprint
  {http://arxiv.org/abs/1810.02136} {arXiv:1810.02136 [hep-ph]} \BibitemShut
  {NoStop}%
\bibitem [{\citenamefont {Chen}\ \emph {et~al.}(2021)\citenamefont {Chen},
  \citenamefont {Zhang}, \citenamefont {Li}, \citenamefont {Hou},\ and\
  \citenamefont {Huang}}]{Chen:2020ath}%
  \BibitemOpen
  \bibfield  {author} {\bibinfo {author} {\bibfnamefont {X.}~\bibnamefont
  {Chen}}, \bibinfo {author} {\bibfnamefont {L.}~\bibnamefont {Zhang}},
  \bibinfo {author} {\bibfnamefont {D.}~\bibnamefont {Li}}, \bibinfo {author}
  {\bibfnamefont {D.}~\bibnamefont {Hou}}, \ and\ \bibinfo {author}
  {\bibfnamefont {M.}~\bibnamefont {Huang}},\ }\href {\doibase
  10.1007/JHEP07(2021)132} {\bibfield  {journal} {\bibinfo  {journal} {JHEP}\
  }\textbf {\bibinfo {volume} {07}},\ \bibinfo {pages} {132} (\bibinfo {year}
  {2021})},\ \Eprint {http://arxiv.org/abs/2010.14478} {arXiv:2010.14478
  [hep-ph]} \BibitemShut {NoStop}%
\bibitem [{\citenamefont {Chen}\ \emph {et~al.}(2020)\citenamefont {Chen},
  \citenamefont {Li}, \citenamefont {Hou},\ and\ \citenamefont
  {Huang}}]{Chen:2019rez}%
  \BibitemOpen
  \bibfield  {author} {\bibinfo {author} {\bibfnamefont {X.}~\bibnamefont
  {Chen}}, \bibinfo {author} {\bibfnamefont {D.}~\bibnamefont {Li}}, \bibinfo
  {author} {\bibfnamefont {D.}~\bibnamefont {Hou}}, \ and\ \bibinfo {author}
  {\bibfnamefont {M.}~\bibnamefont {Huang}},\ }\href {\doibase
  10.1007/JHEP03(2020)073} {\bibfield  {journal} {\bibinfo  {journal} {JHEP}\
  }\textbf {\bibinfo {volume} {03}},\ \bibinfo {pages} {073} (\bibinfo {year}
  {2020})},\ \Eprint {http://arxiv.org/abs/1908.02000} {arXiv:1908.02000
  [hep-ph]} \BibitemShut {NoStop}%
\bibitem [{\citenamefont {Knaute}\ \emph {et~al.}(2018)\citenamefont {Knaute},
  \citenamefont {Yaresko},\ and\ \citenamefont {K\"ampfer}}]{Knaute:2017opk}%
  \BibitemOpen
  \bibfield  {author} {\bibinfo {author} {\bibfnamefont {J.}~\bibnamefont
  {Knaute}}, \bibinfo {author} {\bibfnamefont {R.}~\bibnamefont {Yaresko}}, \
  and\ \bibinfo {author} {\bibfnamefont {B.}~\bibnamefont {K\"ampfer}},\ }\href
  {\doibase 10.1016/j.physletb.2018.01.053} {\bibfield  {journal} {\bibinfo
  {journal} {Phys. Lett. B}\ }\textbf {\bibinfo {volume} {778}},\ \bibinfo
  {pages} {419} (\bibinfo {year} {2018})},\ \Eprint
  {http://arxiv.org/abs/1702.06731} {arXiv:1702.06731 [hep-ph]} \BibitemShut
  {NoStop}%
\bibitem [{\citenamefont {Grefa}\ \emph {et~al.}(2021)\citenamefont {Grefa},
  \citenamefont {Noronha}, \citenamefont {Noronha-Hostler}, \citenamefont
  {Portillo}, \citenamefont {Ratti},\ and\ \citenamefont
  {Rougemont}}]{Grefa:2021qvt}%
  \BibitemOpen
  \bibfield  {author} {\bibinfo {author} {\bibfnamefont {J.}~\bibnamefont
  {Grefa}}, \bibinfo {author} {\bibfnamefont {J.}~\bibnamefont {Noronha}},
  \bibinfo {author} {\bibfnamefont {J.}~\bibnamefont {Noronha-Hostler}},
  \bibinfo {author} {\bibfnamefont {I.}~\bibnamefont {Portillo}}, \bibinfo
  {author} {\bibfnamefont {C.}~\bibnamefont {Ratti}}, \ and\ \bibinfo {author}
  {\bibfnamefont {R.}~\bibnamefont {Rougemont}},\ }\href {\doibase
  10.1103/PhysRevD.104.034002} {\bibfield  {journal} {\bibinfo  {journal}
  {Phys. Rev. D}\ }\textbf {\bibinfo {volume} {104}},\ \bibinfo {pages}
  {034002} (\bibinfo {year} {2021})},\ \Eprint
  {http://arxiv.org/abs/2102.12042} {arXiv:2102.12042 [nucl-th]} \BibitemShut
  {NoStop}%
\bibitem [{\citenamefont {Cai}\ \emph {et~al.}(2022{\natexlab{b}})\citenamefont
  {Cai}, \citenamefont {He}, \citenamefont {Li},\ and\ \citenamefont
  {Wang}}]{Cai:2022omk}%
  \BibitemOpen
  \bibfield  {author} {\bibinfo {author} {\bibfnamefont {R.-G.}\ \bibnamefont
  {Cai}}, \bibinfo {author} {\bibfnamefont {S.}~\bibnamefont {He}}, \bibinfo
  {author} {\bibfnamefont {L.}~\bibnamefont {Li}}, \ and\ \bibinfo {author}
  {\bibfnamefont {Y.-X.}\ \bibnamefont {Wang}},\ }\href {\doibase
  10.1103/PhysRevD.106.L121902} {\bibfield  {journal} {\bibinfo  {journal}
  {Phys. Rev. D}\ }\textbf {\bibinfo {volume} {106}},\ \bibinfo {pages}
  {L121902} (\bibinfo {year} {2022}{\natexlab{b}})},\ \Eprint
  {http://arxiv.org/abs/2201.02004} {arXiv:2201.02004 [hep-th]} \BibitemShut
  {NoStop}%
\bibitem [{\citenamefont {Li}\ \emph {et~al.}(2023)\citenamefont {Li},
  \citenamefont {Liang}, \citenamefont {He},\ and\ \citenamefont
  {Li}}]{Li:2023mpv}%
  \BibitemOpen
  \bibfield  {author} {\bibinfo {author} {\bibfnamefont {Z.}~\bibnamefont
  {Li}}, \bibinfo {author} {\bibfnamefont {J.}~\bibnamefont {Liang}}, \bibinfo
  {author} {\bibfnamefont {S.}~\bibnamefont {He}}, \ and\ \bibinfo {author}
  {\bibfnamefont {L.}~\bibnamefont {Li}},\ }\href {\doibase
  10.1103/PhysRevD.108.046008} {\bibfield  {journal} {\bibinfo  {journal}
  {Phys. Rev. D}\ }\textbf {\bibinfo {volume} {108}},\ \bibinfo {pages}
  {046008} (\bibinfo {year} {2023})},\ \Eprint
  {http://arxiv.org/abs/2305.13874} {arXiv:2305.13874 [hep-ph]} \BibitemShut
  {NoStop}%
\bibitem [{\citenamefont {Rougemont}\ \emph {et~al.}(2024)\citenamefont
  {Rougemont}, \citenamefont {Grefa}, \citenamefont {Hippert}, \citenamefont
  {Noronha}, \citenamefont {Noronha-Hostler}, \citenamefont {Portillo},\ and\
  \citenamefont {Ratti}}]{Rougemont:2023gfz}%
  \BibitemOpen
  \bibfield  {author} {\bibinfo {author} {\bibfnamefont {R.}~\bibnamefont
  {Rougemont}}, \bibinfo {author} {\bibfnamefont {J.}~\bibnamefont {Grefa}},
  \bibinfo {author} {\bibfnamefont {M.}~\bibnamefont {Hippert}}, \bibinfo
  {author} {\bibfnamefont {J.}~\bibnamefont {Noronha}}, \bibinfo {author}
  {\bibfnamefont {J.}~\bibnamefont {Noronha-Hostler}}, \bibinfo {author}
  {\bibfnamefont {I.}~\bibnamefont {Portillo}}, \ and\ \bibinfo {author}
  {\bibfnamefont {C.}~\bibnamefont {Ratti}},\ }\href {\doibase
  10.1016/j.ppnp.2023.104093} {\bibfield  {journal} {\bibinfo  {journal} {Prog.
  Part. Nucl. Phys.}\ }\textbf {\bibinfo {volume} {135}},\ \bibinfo {pages}
  {104093} (\bibinfo {year} {2024})},\ \Eprint
  {http://arxiv.org/abs/2307.03885} {arXiv:2307.03885 [nucl-th]} \BibitemShut
  {NoStop}%
\bibitem [{\citenamefont {Zhao}\ \emph {et~al.}(2024)\citenamefont {Zhao},
  \citenamefont {He}, \citenamefont {Hou}, \citenamefont {Li},\ and\
  \citenamefont {Li}}]{Zhao:2023gur}%
  \BibitemOpen
  \bibfield  {author} {\bibinfo {author} {\bibfnamefont {Y.-Q.}\ \bibnamefont
  {Zhao}}, \bibinfo {author} {\bibfnamefont {S.}~\bibnamefont {He}}, \bibinfo
  {author} {\bibfnamefont {D.}~\bibnamefont {Hou}}, \bibinfo {author}
  {\bibfnamefont {L.}~\bibnamefont {Li}}, \ and\ \bibinfo {author}
  {\bibfnamefont {Z.}~\bibnamefont {Li}},\ }\href {\doibase
  10.1103/PhysRevD.109.086015} {\bibfield  {journal} {\bibinfo  {journal}
  {Phys. Rev. D}\ }\textbf {\bibinfo {volume} {109}},\ \bibinfo {pages}
  {086015} (\bibinfo {year} {2024})},\ \Eprint
  {http://arxiv.org/abs/2310.13432} {arXiv:2310.13432 [hep-ph]} \BibitemShut
  {NoStop}%
\bibitem [{\citenamefont {Fu}\ \emph {et~al.}(2024)\citenamefont {Fu},
  \citenamefont {He}, \citenamefont {Li},\ and\ \citenamefont
  {Li}}]{Fu:2024wkn}%
  \BibitemOpen
  \bibfield  {author} {\bibinfo {author} {\bibfnamefont {Q.}~\bibnamefont
  {Fu}}, \bibinfo {author} {\bibfnamefont {S.}~\bibnamefont {He}}, \bibinfo
  {author} {\bibfnamefont {L.}~\bibnamefont {Li}}, \ and\ \bibinfo {author}
  {\bibfnamefont {Z.}~\bibnamefont {Li}},\ }\href@noop {} {\  (\bibinfo {year}
  {2024})},\ \Eprint {http://arxiv.org/abs/2404.12109} {arXiv:2404.12109
  [hep-ph]} \BibitemShut {NoStop}%
\bibitem [{\citenamefont {Jokela}\ \emph {et~al.}(2024)\citenamefont {Jokela},
  \citenamefont {J\"arvinen},\ and\ \citenamefont {Piispa}}]{Jokela:2024xgz}%
  \BibitemOpen
  \bibfield  {author} {\bibinfo {author} {\bibfnamefont {N.}~\bibnamefont
  {Jokela}}, \bibinfo {author} {\bibfnamefont {M.}~\bibnamefont {J\"arvinen}},
  \ and\ \bibinfo {author} {\bibfnamefont {A.}~\bibnamefont {Piispa}},\
  }\href@noop {} {\  (\bibinfo {year} {2024})},\ \Eprint
  {http://arxiv.org/abs/2405.02394} {arXiv:2405.02394 [hep-th]} \BibitemShut
  {NoStop}%
\bibitem [{\citenamefont {Cai}\ \emph {et~al.}(2024)\citenamefont {Cai},
  \citenamefont {He}, \citenamefont {Li},\ and\ \citenamefont
  {Zeng}}]{Cai:2024eqa}%
  \BibitemOpen
  \bibfield  {author} {\bibinfo {author} {\bibfnamefont {R.-G.}\ \bibnamefont
  {Cai}}, \bibinfo {author} {\bibfnamefont {S.}~\bibnamefont {He}}, \bibinfo
  {author} {\bibfnamefont {L.}~\bibnamefont {Li}}, \ and\ \bibinfo {author}
  {\bibfnamefont {H.-A.}\ \bibnamefont {Zeng}},\ }\href@noop {} {\  (\bibinfo
  {year} {2024})},\ \Eprint {http://arxiv.org/abs/2406.12772} {arXiv:2406.12772
  [hep-th]} \BibitemShut {NoStop}%
\bibitem [{\citenamefont {Sato}\ and\ \citenamefont
  {Yoshida}(2013{\natexlab{d}})}]{Sato:2013iua}%
  \BibitemOpen
  \bibfield  {author} {\bibinfo {author} {\bibfnamefont {Y.}~\bibnamefont
  {Sato}}\ and\ \bibinfo {author} {\bibfnamefont {K.}~\bibnamefont {Yoshida}},\
  }\href {\doibase 10.1007/JHEP08(2013)002} {\bibfield  {journal} {\bibinfo
  {journal} {JHEP}\ }\textbf {\bibinfo {volume} {08}},\ \bibinfo {pages} {002}
  (\bibinfo {year} {2013}{\natexlab{d}})},\ \Eprint
  {http://arxiv.org/abs/1304.7917} {arXiv:1304.7917 [hep-th]} \BibitemShut
  {NoStop}%
\bibitem [{\citenamefont {Zhang}\ \emph {et~al.}(2018)\citenamefont {Zhang},
  \citenamefont {Hou},\ and\ \citenamefont {Li}}]{Zhang:2018hfd}%
  \BibitemOpen
  \bibfield  {author} {\bibinfo {author} {\bibfnamefont {L.}~\bibnamefont
  {Zhang}}, \bibinfo {author} {\bibfnamefont {D.-F.}\ \bibnamefont {Hou}}, \
  and\ \bibinfo {author} {\bibfnamefont {J.}~\bibnamefont {Li}},\ }\href
  {\doibase 10.1140/epja/i2018-12524-4} {\bibfield  {journal} {\bibinfo
  {journal} {Eur. Phys. J. A}\ }\textbf {\bibinfo {volume} {54}},\ \bibinfo
  {pages} {94} (\bibinfo {year} {2018})}\BibitemShut {NoStop}%
\bibitem [{\citenamefont {Zhu}\ \emph {et~al.}(2020)\citenamefont {Zhu},
  \citenamefont {Hou},\ and\ \citenamefont {Chen}}]{Zhu:2019igg}%
  \BibitemOpen
  \bibfield  {author} {\bibinfo {author} {\bibfnamefont {Z.-R.}\ \bibnamefont
  {Zhu}}, \bibinfo {author} {\bibfnamefont {D.-f.}\ \bibnamefont {Hou}}, \ and\
  \bibinfo {author} {\bibfnamefont {X.}~\bibnamefont {Chen}},\ }\href {\doibase
  10.1140/epjc/s10052-020-8110-8} {\bibfield  {journal} {\bibinfo  {journal}
  {Eur. Phys. J. C}\ }\textbf {\bibinfo {volume} {80}},\ \bibinfo {pages} {550}
  (\bibinfo {year} {2020})},\ \Eprint {http://arxiv.org/abs/1912.05806}
  {arXiv:1912.05806 [hep-ph]} \BibitemShut {NoStop}%
\bibitem [{\citenamefont {Gelis}\ and\ \citenamefont
  {Tanji}(2016)}]{Gelis:2015kya}%
  \BibitemOpen
  \bibfield  {author} {\bibinfo {author} {\bibfnamefont {F.}~\bibnamefont
  {Gelis}}\ and\ \bibinfo {author} {\bibfnamefont {N.}~\bibnamefont {Tanji}},\
  }\href {\doibase 10.1016/j.ppnp.2015.11.001} {\bibfield  {journal} {\bibinfo
  {journal} {Prog. Part. Nucl. Phys.}\ }\textbf {\bibinfo {volume} {87}},\
  \bibinfo {pages} {1} (\bibinfo {year} {2016})},\ \Eprint
  {http://arxiv.org/abs/1510.05451} {arXiv:1510.05451 [hep-ph]} \BibitemShut
  {NoStop}%
\end{thebibliography}%

\end{document}